\documentclass[%
 reprint,
amsmath,
amssymb,
aps,
floatfix,
superscriptaddress,
]{revtex4-2}

\usepackage{graphicx}
\usepackage{dcolumn}
\usepackage{bm}
\usepackage{nccmath}
\usepackage{amsmath}
\usepackage{bm}
\usepackage{empheq}
\usepackage{dcolumn}
\usepackage{mathtools}
\usepackage{tensor}
\usepackage{bm}
\usepackage{subfigure}
\usepackage{comment}
\providecommand{\abs}[1]{\lvert#1\rvert} \providecommand{\norm}[1]{\lVert#1\rVert}

\usepackage{orcidlink}

\usepackage[normalem]{ulem}
\usepackage{mathrsfs}
\begin{document}

\preprint{APS/123-QED}

\title{Scalar-tensor theories in the Lyra geometry: Invariance under local transformations of length units and the Jordan-Einstein frame conundrum}%

\author{E. C. Valadão \orcidlink{0009-0004-1172-7464}}%
\email{eduardo.valadao@edu.ufes.br}
\affiliation{%
 Centro Brasileiro de Pesquisas Físicas (CBPF), Rua Dr.~Xavier Sigaud 150, CEP 22290-180, Rio de Janeiro, RJ, Brazil \\
}
\affiliation{%
 PPGCosmo, Centro de Ciências Exatas, Universidade Federal do Espírito Santo (UFES),
 CEP 29075-910, Vitória, ES, Brazil \\
}

\author{Felipe Sobrero \orcidlink{0009-0009-6497-6612}}
\email{felipesobrero@cbpf.br}
\affiliation{%
 Centro Brasileiro de Pesquisas Físicas (CBPF), Rua Dr.~Xavier Sigaud 150, CEP 22290-180, Rio de Janeiro, RJ, Brazil \\
}

\author{Santiago Esteban Perez Bergliaffa \orcidlink{0000-0002-2525-0386}}%
\affiliation{%
Departamento de Física Teórica, Instituto de Física,
Universidade do Estado do Rio de Janeiro (UERJ),
Rua São Francisco Xavier 524, Maracanã
CEP 20550-013, Rio de Janeiro, RJ, Brazil \\
}%

\begin{abstract}
The Lyra geometry provides an interesting approach to develop purely geometrical scalar-tensor theories due to the natural presence of the Lyra scale function. This paper explores further the scale function source term to construct a theory on Lyra manifolds which contains proper generalizations of both Brans-Dicke gravity and the Einstein-Gauss-Bonnet scalar-tensor theory. It is shown that the symmetry group of gravitational theories on the Lyra geometry comprises not only coordinate transformations but also local transformations of length units, so that the Lyra function is a conformal factor which locally fixes the unit of length. The Lyra geometry is thus a generalization of Riemannian geometry which includes spacetime-dependent length units. By performing a Lyra transformation to a frame in which the unit of length is globally fixed, it is shown that General Relativity (GR) is obtained from the Lyra Scalar-Tensor Theory (LyST). Through the same procedure, even in the presence of matter fields, it is found that Brans-Dicke gravity and the Einstein-Gauss-Bonnet scalar-tensor theory are obtained from their Lyra counterparts. It is argued that this approach is consistent with the Mach-Dicke principle, since the strength of gravity in Brans-Dicke-Lyra is controlled by the scale function. It might be possible that any known scalar-tensor theory can be naturally geometrized by considering a particular Lyra frame, for which the scalar field is the function which locally controls the unit of length. The Jordan-Einstein frame conundrum is also assessed from the perspective of Lyra transformations, it is shown that the Lyra geometry makes explicit that the two frames are only different representations of the same theory, so that in the Einstein frame the unit of length varies locally. The Lyra formalism is then shown to be better suited for exploring scalar-tensor gravity, since in its well-defined structure the conservation of the energy-momentum tensor and geodesic motion are assured in the Einstein frame.
\end{abstract}

\maketitle

\section{Introduction}
\label{sec:1} 

The Lyra geometry, proposed by Gerhard Lyra in 1951 \cite{Lyra1951}, is an extension of Riemannian geometry constructed with the purpose of incorporating general gauge invariance \cite{Lyra1951}. An important aspect of its formalism is that a reference system comprises not only a coordinate system but also a gauge function \cite{Lyra1951, Scheibe_1952, Sen_1971}. Currently referred to as the Lyra scale function \cite{Cuzinatto2021, Sobrero2024}, the gauge function is introduced in the structureless manifold. Hence, it is \textit{a priori} independent of the metric structure \cite{Lyra1951, Sen_1971}, in spite of the fact that it bears great similarities with the Weyl geometry \cite{Weyl1918a, Weyl_1918, Weyl_1922, Romero_2019}. As a consequence, the resulting metric-compatible connection of the Lyra geometry is very similar to the connection of Weyl integrable spacetimes (WISTs) \cite{Novello_1983, Romero_2019}, so that the length of vectors is also path-independent under parallel transport, while maintaining a vanishing nonmetricity tensor \cite{Sobrero2024}. 

Since its inception, however, Lyra manifolds have not been thoroughly explored in the context of gravitational theories. It was not until 2021 that the Lyra scale function was used as a gravitational field alongside the metric, when the Lyra Scalar-Tensor theory (LyST) was introduced in \cite{Cuzinatto2021}. In previous works, an auxiliary vector field was used instead in the variational principle \cite{Sen_1957, Sen_1971, Jeavons_1975}. Nevertheless, since the vector field employed is naturally related to the Lyra scale function, the authors of \cite{Cuzinatto2021} proposed that this method is not strictly necessary to obtain a geometrical Lyra scalar-tensor theory, since the use of the Lyra function itself as a fundamental field is a much more simple procedure.  

It follows then that Lyra manifolds offer an approach for which purely geometric four-dimensional scalar-tensor theories can be naturally developed, such that the scalar field has a simple geometric origin within the Lyra framework. This theoretical perspective is of particular importance due to the success of General Relativity (GR) in the description of many phenomena at the scale of the Solar System \cite{Will2014a, Will2014b}, compact objects \cite{Taylor1989, Abbott2019a, Akiyama2019a} and Cosmology \cite{Planck2020}, as one of its main conceptual features is the description of gravity via the geometrization of spacetime. 

Therefore, one of the purposes of this work is to construct a scalar-tensor theory in which the scalar has a 4D geometrical nature. In particular, this goal is attained by formulating a Lagrangian density on Lyra manifolds which generalizes Brans-Dicke gravity \cite{Brans1961} and which also contains the Gauss-Bonnet term \cite{Lanczos1938} of the Lyra geometry. However, as far as the authors are aware, the Lyra function source term was not yet properly explored in the literature. So to construct the theory mentioned above, the scale function source term is thoroughly analyzed for the general case of an imperfect fluid. It is shown that the form of the resulting source term is an essential element of the Lyra geometrization procedure. 

Regarding the Gauss-Bonnet term, it is well-known that this scalar does not modify GR field equations when added to the Einstein-Hilbert Lagrangian density \cite{Lanczos1938}. This curvature invariant yields an expression proportional to the Euler characteristic of the manifold \cite{Tian2016}, for which the variation vanishes. As a consequence, alternative approaches are necessary if Gauss-Bonnet contributions are to appear in four-dimensional field equations \cite{Fernandes2022, Sotiriou2007}. To achieve such objective, it is necessary a metric-affine geometry \cite{Haghani2015, Jimenez2014}, a function which is not linear in the Gauss-Bonnet scalar \cite{Nojiri2005a, Myrzakulov2011, Anjos2024} or even consider a higher-dimensional manifold and take the limit to four dimensions \cite{Glavan2020, Fernandes2022}. Nonetheless, the most used approach is to consider a dynamical scalar field non-minimally coupled to this topological invariant \cite{Boulware1985, Antoniadis1994, Rizos1994, Nojiri2005b}.

Scalar-tensor theories with the Gauss-Bonnet invariant possess a very interesting phenomenology. They have scalar-hairy black hole solutions \cite{Sotiriou2014, Babichev2024, Saravani2019} and describe spontaneous scalarization \cite{Doneva2018, Canate2020, Silva2018, Cunha2019}. In Cosmology, this class of theories can account for inflation \cite{Jiang2013, Koh2014, Laurentis2015, Chakraborty2018, Odintsov2018}, for late-time acceleration \cite{Vernov2021, Tsujikawa2007, Nojiri2017, Martino2020, Benetti2018, Sadjadi2024, Odintsov2021, Guendelman2019, Jimenez2019} and can also prevent the formation of singularities \cite{Antoniadis1994, Rizos1994, Kanti1999, Odintsov2020b}. But, in essence, the main advantage of Gauss-Bonnet gravity is that, although it is a second-order term in Riemann curvature objects, the resulting contribution to the field equations has no third- or fourth-order derivatives of the metric \cite{Fernandes2022}, thus circumventing Ostrogradsky instabilities  \cite{Fernandes2022}. 

However, although inspired by quantum corrections that appear in the effective action of some models in heterotic string theory \cite{Fernandes2022, Boulware1985, Antoniadis1994}, the scalar-tensor Einstein-Gauss-Bonnet theories are not completely geometric theories. There are only particular cases for which the scalar has its origin related to string properties or to exotic compactified higher dimensions \cite{Boulware1985, Antoniadis1994}. As for the Brans-Dicke theory, a similar situation occurs, since it can be obtained from the Kaluza-Klein theory \cite{Faraoni2004}. Nevertheless, in general, and in the mentioned cases, the scalar still does not possess a completely four-dimensional geometrical nature, so that the Lyra geometry can thus provide a solution to this issue.  

Another essential topic that remains largely unexplored in the literature is the change between the Jordan and Einstein frame in the context of scalar-tensor theories on Lyra manifolds. Although the debate of the equivalence of these frames is plagued by confusion (see \cite{Faraoni2004, Quiros2019} for the various perspectives), the controversy at the classical level is settled if one adopts the original interpretation made by Robert H. Dicke in 1962 \cite{Dicke1962, Faraoni2004, Faraoni2007}. In Dicke's original work \cite{Dicke1962}, it is shown that a locally-varying system of units can be adopted to study Brans-Dicke gravity without altering the physics \cite{Dicke1962, Faraoni2004, Faraoni2007, Deruelle2011, Chiba2013, Morris2014, Postma2014, Quiros2012, Quiros2018}. Nonetheless, as noted in \cite{Faraoni2004, Faraoni2007}, this work has been widely misinterpreted in the literature, leading to the view that one of the frames must be unphysical \cite{Faraoni2004}.  

By considering the interpretation of Dicke, this paper studies the Jordan and Einstein frame conundrum \cite{Faraoni2004, Faraoni2007, Quiros2012, Quiros2019, Velasquez2023} in Lyra manifolds. As a matter of fact, the Lyra geometry is perfectly appropriate for the implementation of Dicke's ideas \cite{Dicke1962}, since this present work shows that Lyra transformations are directly related to transformations between systems of units which are spacetime-dependent. This perspective is the core aspect of the paper, since it also leads to a straightforward relation between the geometrized theory constructed and the usual Brans-Dicke theory and scalar-tensor Gauss-Bonnet gravity of Riemannian manifolds. In this sense, it is important to note that many attempts have been made as to create a purely 4D geometrical Brans-Dicke theory, e.g. \cite{Sen_1971, Soleng1988, Punzi2008, Almeida2014}, but none of them can completely recover the original theory. Nonetheless, as shown here, the Lyra geometry along with the locally varying length unit interpretation can perfectly solve this issue.

The paper is organized as follows. In section \ref{sec2}, based on \cite{Sobrero2024, Lyra1951, Sen_1971, Cuzinatto2021, Scheibe_1952}, a brief review of the Lyra geometry and its metric and affine structures is presented. In section \ref{sec3}, a review of the LyST theory is shown along with the definition of the scale function matter source term \cite{Cuzinatto2021}. As for section \ref{sec4}, a generalization of the approach in \cite{Mendoza2021, Mendoza2023} is used to find the Lyra scale function source term for a generic imperfect fluid. The Lyra theory generalizing Brans-Dicke theory and the scalar-tensor Gauss-Bonnet gravity is shown in section \ref{sec6}. Section \ref{sec5} is devoted to the relation between local transformations of length units and Lyra frame transformations, along with its consequences to the Lyra geometric objects. In subsection \ref{sec5.1} it is shown the relation of the LyST theory with GR and in \ref{subsec6.1} the correspondence of the Lyra theory constructed here with its Riemannian counterpart. The Jordan and Einstein frame conundrum from the perspective of the Lyra geometry is presented in section \ref{sec7}. Our final comments are given in section \ref{sec8}.

\section{Lyra geometry}
\label{sec2}

An $n$-dimensional differentiable Lyra manifold is a real set equipped with a maximal atlas that contains a special gauge function, called Lyra scale function. It is defined as $\phi \coloneq \Phi_{k} \circ \mathcal{X}^{-1}_{k}$, in which the chart $\mathcal{X}_{k}$ and the $\mathcal{C^{\infty}}$ scale map $\Phi_{k}$, defined over an open subset $\mathcal{O}_{k}$, characterize a Lyra reference system \cite{Sen_1971, Sobrero2024}.

The fundamental aspect of this geometry is that the canonical basis for an $n$-dimensional tangent vector space $\mathcal{T}_{p}$ at a point $p$ of the manifold is constructed as
\begin{equation}
\label{eq1}
\mathbf{e}_{\mu}f \coloneq \frac{1}{\phi(x)} \frac{\partial (f \circ \mathcal{X}^{-1}_{k})}{\partial x^{\mu}}\Biggr|_{\substack{x(p)}},
\end{equation}
in which $f$ is a $\mathcal{C^{\infty}}$ function and $x^{\mu}$ are the coordinates in the chart $\mathcal{X}_{k}$. An immediate 
consequence 
of this definition is that the Lyra geometry possesses a canonical basis with noncommutative elements \cite{Cuzinatto2021}. 

Due to definition \eqref{eq1}, the components and basis elements of a vector $\mathbf{v} = v^{\mu} \mathbf{e}_{\mu}$ must transform as
\begin{equation}
\label{eq2}
\bar{\mathbf{e}}_{\mu} = \frac{\phi (x)}{\bar{\phi} (\bar{x})} \frac{\partial x^{\nu}}{\partial \bar{x}^{\mu}} \mathbf{e}_{\nu} \ \ \text{and} \ \ \bar{v}^{\mu} = \frac{\bar{\phi} (\bar{x})}{\phi (x)} \frac{\partial \bar{x}^{\mu}}{\partial x^{\nu}} v^{\nu},
\end{equation}
upon the change of reference system
$(\mathcal{O}; \mathcal{X}, \Phi) \rightarrow (\bar{\mathcal{O}}; \bar{\mathcal{X}}, \bar{\Phi})$ \cite{Sobrero2024}. Thus, a change between different Lyra reference systems is both a coordinate and a scale transformation. Another important consequence is that, since a tangent vector to a smooth curve $\gamma(\lambda)$ is straightforwardly associated with a directional derivative, the components of a Lyra tangent vector are then defined by \cite{Sobrero2024} 
\begin{equation}
\label{eq3}
v^{\mu} = \phi(x) \frac{d x^{\mu}}{d \lambda}.
\end{equation}  

As for the dual Lyra vector space $\mathcal{T}^{*}_{p}$, the natural basis definition is given by \cite{Cuzinatto2021}
\begin{equation}
\label{eq3.1}
\bm{\theta}^{\mu} = \phi(x) \mathbf{d}x^{\mu}, 
\end{equation}
which is a result of the orthonormality condition $\bm{\theta}^{\mu} \circ \mathbf{e}_{\nu} = \tensor{\delta}{^\mu_\nu}$ \cite{Sobrero2024}. Thus, as in Riemannian geometry, a dual vector basis transform as the vector components in \eqref{eq2}. Hence, its components must transform as the vector basis.

Having defined vectors and their duals, a Lyra tensor is then defined as a multilinear map that takes $k$ elements of $\mathcal{T}^{*}_{p}$ and $l$ vectors of $\mathcal{T}_{p}$ to the real numbers, so that a generic tensor $T$ is defined as
\begin{equation}
\label{eq3.2}
T = T^{\mu_{1}...\mu_{k}}_{\ \ \ \ \ \ \ \ \nu_{1}...\nu_{l}} \bm{e}_{\mu_{1}} \otimes ... \otimes \bm{e}_{\mu_{k}} \otimes \bm{\theta}^{\nu_{1}} \otimes ... \otimes \bm{\theta}^{\nu_{l}}.
\end{equation}
As a consequence, the Lyra transformation law for the components of tensors is given by \cite{Sobrero2024}
\begin{equation}
\label{eq4}
\hspace*{-0.1cm}
\bar{T}^{\mu_{1}...\mu_{k}}_{\ \ \ \ \ \ \ \ \nu_{1}...\nu_{l}}=\bigg(\frac{\bar{\phi} (\bar{x})}{\phi (x)}\bigg)^{k-l} \frac{\partial \bar{x}^{\mu_{1}}}{\partial x^{\eta_{1}}} ... \frac{\partial x^{\xi_{l}}}{\partial \bar{x}^{\nu_{l}}} T^{\eta_{1}...\eta_{k}}_{\ \ \ \ \ \ \ \xi_{1}...\xi_{l}},
\end{equation}
such that the contraction, sum, scalar product and tensorial product of these objects are defined as in Riemannian manifolds \cite{Sen_1971, Cuzinatto2021}.

\subsection{The metric structure} \label{subsec2.1}

The metric tensor is a smooth bilinear symmetric nondegenerate map which allows the definition of spatio-temporal intervals and the notion of causality. It defines the inner product as $\mathbf{g}(\mathbf{v}, \mathbf{u}) = \tensor{g}{_\mu_\nu}v^{\mu}u^{\nu}$, such that $\tensor{g}{_\mu_\nu} \coloneq \mathbf{g}(\mathbf{e}_{\mu}, \mathbf{e}_{\nu})$ are the metric components. As in Riemannian manifolds, this mathematical entity enables the canonical identification of a vector space and its dual by establishing $v_{\mu} \coloneq \mathbf{g}(\mathbf{v}, \mathbf{e}_{\mu}) = \tensor{g}{_\mu_\nu} v^{\nu}$. In the same manner, the inverse metric is defined via $\tensor{g}{^\mu^\alpha} \tensor{g}{_\alpha_\nu} = \tensor{\delta}{^\mu_\nu}$ and the inner product is not necessarily positive definite. 

An interesting aspect though is that the length of vectors $\norm{\mathbf{v}}^{2} \coloneq \mathbf{g}(\mathbf{v}, \mathbf{v})$ displays the Lyra scale function as a multiplicative factor in the corresponding  Riemannian expression, which leads to the mentioned similarities with Weyl Integrable Spacetimes \cite{Sobrero2024}. As a consequence, a normalization such as $v^{\mu} v_{\mu} = c^{2}$ is naturally possible in a metric compatible Lyra manifold. A further important result of this definition of length is the line element of the Lyra geometry, which is given by
\begin{equation}
\label{eq5} 
ds^{2} = \phi^{2} \tensor{g}{_\mu_\nu} dx^{\mu} dx^{\nu},
\end{equation} 
such that the presence of $\phi^{2}$, originating from the dual vector basis definition \eqref{eq3.1}, makes this expression invariant under Lyra reference system transformations \cite{Cuzinatto2021}. 

The equation for the geodesics is obtained as usual, by demanding that such a curve extremizes the interval connecting two given events, which leads to \cite{Sobrero2024}
\begin{empheq}{align}
\label{eq6}
  &\frac{d^{2}x^{\gamma}}{ds^{2}} + \bigg\{\genfrac{}{}{0pt}{0}{\gamma}{\mu \nu}\bigg\}\frac{dx^{\mu}}{ds}\frac{dx^{\nu}}{ds} \nonumber \\
  &+ \frac{1}{\phi} (\tensor{\delta}{^\gamma_\nu} \partial_{\mu} \phi + \tensor{\delta}{^\gamma_\mu} \partial_{\nu} \phi - \tensor{g}{_\mu_\nu} \partial^{\gamma} \phi) \frac{dx^{\mu}}{ds}\frac{dx^{\nu}}{ds} = 0. 
\end{empheq}
Another crucial modification that appears in the metric structure, determined by considering the transformation of the metric determinant $g$ \cite{Cuzinatto2021, Sobrero2024}, is the Lyra volume $n$-form $dV = \phi^{n} \sqrt{\abs{g}} d^{n}x$.

\subsection{The affine structure} \label{subsec2.2}

The map $\bm{\nabla}\text{:} \ \mathcal{T}_{p} \times \mathcal{T}_{p} \rightarrow \mathcal{T}_{p}$, which takes two vectors $\mathbf{u}$ and $\mathbf{v}$ to the object $\bm{\nabla}_{\mathbf{u}} \mathbf{v}$, enables the connection of different points of the manifold. The structure added by this entity provides the description of inertial motion. Using the canonical Lyra basis \eqref{eq1} and the requirement of linearity \cite{Sobrero2024}, it follows that
\begin{equation}
\label{eq7}
\bm{\nabla}_{\mathbf{u}}\mathbf{v} = u^{\nu} \big(\nabla_{\nu}v^{\alpha}\big) \mathbf{e}_{\alpha} = u^{\nu} \bigg(\frac{1}{\phi}\partial_{\nu}v^{\alpha} + \Gamma^{\alpha}_{\mu\nu}v^{\mu}\bigg) \mathbf{e}_{\alpha},
\end{equation}
in which $(\nabla_{\nu}v^{\alpha})\mathbf{e}_{\alpha} \coloneq \bm{\nabla}_{\mathbf{e}_{\nu}}\mathbf{v}$ is called Lyra covariant derivative and $\Gamma^{\alpha}_{\mu\nu} \mathbf{e}_{\alpha} \coloneq \bm{\nabla}_{\mathbf{e}_{\nu}}\mathbf{e}_{\mu}$ is the affine connection. 

The covariant derivative of a scalar function $f$ can be thus naturally defined as $\nabla_{\mu}f = \phi^{-1} \partial_{\mu} f$. As a result, using this property on the contraction of a contravariant vector with a covariant one, it follows that there is a map for dual vectors, given by $\nabla_{\mu} \omega_{\nu} = \phi^{-1}\partial_{\mu} \omega_{\nu} - \Gamma^{\alpha}_{\nu\mu} \omega_{\alpha}$ \cite{Sobrero2024}. With this result and expression \eqref{eq7}, the covariant derivative of a generic Lyra tensor is thus found as in Riemann manifolds \cite{Cuzinatto2021}.

The affine connection in expression \eqref{eq7} brings a curve-dependent notion of parallelism between vectors on different points of the manifold. It also leads to a special type of curve whose tangent vector is parallel transported along it, namely an autoparallel curve, mathematically described by  $\bm{\nabla}_{\mathbf{v}} \mathbf{v} = \bm{0}$, which yields
\begin{equation}
\label{eq8}
\frac{d^{2}x^{\alpha}}{d\lambda^{2}} + \big(\phi \Gamma^{\alpha}_{\mu\nu} + \tensor{\delta}{^\alpha_\mu} \nabla_{\nu}\phi \big) \frac{dx^{\mu}}{d\lambda}\frac{dx^{\nu}}{d\lambda} = 0.
\end{equation}

Having introduced the connection, the Riemann curvature tensor on the Lyra geometry is obtained through the non-commutativity of successive covariant differentiations of a vector field. Using the covariant derivative defined in \eqref{eq7}, it follows that \cite{Sobrero2024}
\begin{equation}
\label{eq9}
\tensor{R}{^\rho_\mu_\gamma_\alpha} = \frac{2}{\phi^{2}}\partial_{[\gamma} \big(\phi \Gamma^{\rho}_{|\mu|\alpha]}\big) + 2\Gamma^{\rho}_{\sigma[\gamma} \Gamma^{\sigma}_{|\mu|\alpha]},
\end{equation}
in which the vertical bar symbols are excluding the index inside them of the antisymmetrization procedure represented by the brackets. The same approach also yields the Lyra torsion tensor
\begin{equation}
\label{eq10}
\tensor{\tau}{^\rho_\gamma_\alpha} = \Gamma^{\rho}_{\alpha\gamma} - \Gamma^{\rho}_{\gamma\alpha} + \frac{1}{\phi} \big(\tensor{\delta}{^\rho_\alpha} \nabla_{\gamma} \phi -  \tensor{\delta}{^\rho_\gamma} \nabla_{\alpha} \phi\big),
\end{equation}
where the last term is equal to the basis non-commutativity tensor shown in \cite{Cuzinatto2021}. Note that the symbol $\nabla_{\mu}$ refers to the Lyra covariant derivative.

The non-metricity tensor $\tensor{Q}{_\alpha_\mu_\nu} \coloneq - \nabla_{\alpha} \tensor{g}{_\mu_\nu}$ relates the affine and metric structures of a manifold. For a metric compatible ($\tensor{Q}{_\alpha_\mu_\nu} = 0$) and torsionless ($\tensor{\tau}{^\rho_\gamma_\alpha} = 0$) manifold, the relation $-\tensor{Q}{_\alpha_\mu_\nu} + \tensor{Q}{_\mu_\nu_\alpha} + \tensor{Q}{_\nu_\alpha_\mu} = 0$ leads to the Lyra connection
\begin{equation}
\label{eq11}
\Gamma^{\gamma}_{\mu\nu} = \frac{1}{\phi} \bigg\{\genfrac{}{}{0pt}{0}{\gamma}{\mu \nu}\bigg\} + \frac{1}{\phi}\big(\tensor{\delta}{^\gamma_\nu} \nabla_{\mu} \phi - \tensor{g}{_\mu_\nu}\nabla^{\gamma} \phi \big).
\end{equation}
In this case, the autoparallel equation \eqref{eq8} coincides with the geodesic one in \eqref{eq6}.

\section{The Lyra Scalar-Tensor Theory} \label{sec3}

In reference \cite{Cuzinatto2021}, along with the metric tensor, the Lyra scale function is considered a fundamental field in the description of gravity. It is important to separate the contributions of each field when dealing with the curvature tensors. For $\tensor{\tau}{^\rho_\gamma_\alpha} = \tensor{Q}{_\alpha_\mu_\nu} = 0$, the Lyra-Riemann tensor \eqref{eq9}, for example, can be written as
\begin{eqnarray}
\label{eq11.1}
  \tensor{R}{^\rho_\mu_\gamma_\alpha} &=& \frac{1}{\phi^{2}} \tensor{\mathcal{R}}{^\rho_\mu_\gamma_\alpha} \ + \ \frac{1}{\phi^{2}} \tensor{g}{^\rho^\beta} \tensor{g}{_\mu_\alpha_\beta_\gamma} \nabla_{\sigma} \phi \nabla^{\sigma} \phi \nonumber \\
  && + \frac{1}{\phi} \tensor{g}{^\rho^\beta} (\tensor{g}{_\mu_\nu} \tensor{g}{_\beta_\alpha_\sigma_\gamma} - \tensor{g}{_\beta_\nu} \tensor{g}{_\mu_\alpha_\sigma_\gamma}) \nabla^{\sigma} \nabla^{\nu} \phi, 
\end{eqnarray}
such that $\nabla_{\mu}$ is the Lyra covariant derivative, $\tensor{g}{_\mu_\alpha_\beta_\gamma} \coloneq \tensor{g}{_\mu_\alpha} \tensor{g}{_\beta_\gamma} - \tensor{g}{_\mu_\gamma} \tensor{g}{_\beta_\alpha}$ and $\tensor{\mathcal{R}}{^\rho_\mu_\gamma_\alpha}$ is the usual curvature tensor of Riemannian geometry with respect to the Christoffel symbols. As a consequence, the Lyra version of the Ricci tensor is given by
\begin{eqnarray}
\label{eq12}
  \tensor{R}{_\mu_\nu} &=& \frac{1}{\phi^{2}} \tensor{\mathcal{R}}{_\mu_\nu} \ - \ \frac{2}{\phi} \nabla_{\nu} \nabla_{\mu} \phi \ - \ \frac{1}{\phi} \tensor{g}{_\mu_\nu} \Box \phi \nonumber \\
  && + \frac{3}{\phi^{2}} \tensor{g}{_\mu_\nu} {\nabla^{\rho} \phi} {\nabla_{\rho} \phi}, 
\end{eqnarray}
in which $\tensor{\mathcal{R}}{_\mu_\nu}$ is defined with the Christoffel symbols and such that $\Box \coloneq \nabla^\rho \nabla_{\rho}$. The Lyra-Ricci tensor is also symmetric, since $\nabla_{[\nu} \nabla_{\mu]} \phi = 0$.  Therefore, the Ricci scalar in a four-dimensional Lyra manifold can be expressed as
\begin{equation}
\label{eq13}
R = \frac{1}{\phi^{2}} \mathcal{R} \ - \ \frac{6}{\phi} \Box \phi \ + \ \frac{12}{\phi^{2}} {\nabla^{\rho} \phi} {\nabla_{\rho} \phi},
\end{equation}
in which $\mathcal{R} \coloneq \tensor{g}{^\mu^\nu} \tensor{\mathcal{R}}{_\mu_\nu}$.

The Lyra Scalar-Tensor Theory (LyST) has the Ricci scalar \eqref{eq13} as its Lagrangian density, namely,
\begin{equation}
\label{eq14.1}
\hspace{-0.2cm}
S[\tensor{g}{_\mu_\nu}, \phi, \psi_{a}] = \int_{\mathcal{M}} \frac{R}{2k} \,\phi^4 \sqrt{-g} \,d^{4}x + S_{m}[\tensor{g}{_\mu_\nu}, \phi, \psi_{a}],
\end{equation}
where $S_{m}$ depends on matter fields $\psi_{a}$ and its first derivatives, as well as on the Lyra field. The LyST field equations are thus given by \cite{Cuzinatto2021}
\begin{eqnarray}
\label{eq14}
  \frac{1}{\phi^{2}} \tensor{\mathcal{G}}{_\mu_\nu} - \ \frac{2}{\phi} \nabla_{\mu} \nabla_{\nu} \phi \ + \ \frac{2}{\phi} \tensor{g}{_\mu_\nu} \Box \phi && \nonumber \\
   - \ \frac{3}{\phi^{2}} \tensor{g}{_\mu_\nu} {\nabla^{\rho} \phi} {\nabla_{\rho} \phi} &=& \kappa \tensor{T}{_\mu_\nu},
\end{eqnarray}
such that $\tensor{\mathcal{G}}{_\mu_\nu} \coloneq \tensor{\mathcal{R}}{_\mu_\nu} - \frac{1}{2} \mathcal{R} \tensor{g}{_\mu_\nu}$ and with the energy-momentum tensor being defined as in GR \cite{Sobrero2024}:
\begin{equation}
\label{eq14.2}
\tensor{T}{_\mu_\nu} \coloneq - \frac{2}{\sqrt{-g}} \frac{\delta (\mathcal{L}_{m} \sqrt{-g})}{\delta \tensor{g}{^\mu^\nu}}.
\end{equation}
As a matter of fact, General Relativity is obtained if $\phi = 1$ \cite{Cuzinatto2021}. Moreover, the variation of $S$ with respect to the matter fields furnishes the equations of motion for the matter, which in general depend on the Lyra field $\phi$. It is important to note that Eq. \eqref{eq14} is obtained considering the divergence theorem of the Lyra geometry, i.e. 
\begin{equation}
\label{eq14.3}
\int_{V} \nabla_{\mu} A^{\mu} \phi^{4} \sqrt{-g} \,d^{4}x = \oint_{\partial V} n_{\mu} A^{\mu} \phi^{3} \sqrt{|\gamma|} \,d^{3}x,
\end{equation}
which uses the Lyra covariant derivative and such that $\gamma$ is the projected metric on $\partial V$.

Finally, variation of the action \eqref{eq14.1} with respect to the Lyra scale function leads, using relation \eqref{eq13}, to $R = \kappa \Omega$, in which the scale function source term is given by
\begin{equation}
\label{eq15}    
\Omega = -4\mathcal{L}_{m} - \phi \frac{\delta \mathcal{L}_{m}}{\delta \phi}.
\end{equation}
Since taking the trace of Eq.\eqref{eq14} leads to $R = - \kappa T$, it follows that $\Omega = -T$. As it will be shown in the next section, the latter is actually an identity, so the result for the variation of the scalar field is actually given by $R = - \kappa T$. Explicitly,
\begin{equation}
\label{eq15.1}
\frac{1}{\phi^{2}} \mathcal{R} \ - \ \frac{6}{\phi} \Box \phi \ + \ \frac{12}{\phi^{2}} {\nabla^{\rho} \phi} {\nabla_{\rho} \phi}=-\kappa T.
\end{equation}

\section{The scale function source} 
\label{sec4}

As shown in section \ref{sec3}, the variation of the action \eqref{eq14.1} with respect to the Lyra field yields $R = \kappa \Omega$, such that $\Omega$ is given by Eq.\eqref{eq15}. In the LyST case, it is trivially found that $\Omega = - T$, where $T = \tensor{g}{^\mu^\nu} \tensor{T}{_\mu_\nu}$ is the trace of the energy-momentum tensor. For Maxwell´s electromagnetic Lagrangian on Lyra manifolds \cite{Sobrero2024} and for a scalar field with $\mathcal{L} = \mathcal{L}(X,\varphi)$, where $X \equiv - \nabla_\mu \varphi \nabla^\mu \varphi / 2$, $\Omega = - T$ holds, as it can be straightforwardly found using Eq.\eqref{eq15}. For theories beyond the LyST, as the one in section \ref{sec6}, $\Omega = - T$ cannot be found by using solely the field equations. Therefore, such an identity must be verified in the most general fluid scenario and regardless of the gravitational Lagrangian used.   

For this purpose, the approach of \cite{Mendoza2021, Mendoza2023} is generalized here to obtain the matter Lagrangian, and $\delta \mathcal{L}_{m}/\delta \phi$, for an imperfect fluid on Lyra manifolds. 

The energy-momentum tensor can thus be written as 
\begin{equation}
\label{eq20}
\tensor{T}{_\mu_\nu} = (\rho + p)v_{\mu}v_{\nu} - p \tensor{g}{_\mu_\nu} + 2 q_{(\mu}v_{\nu)} + \tensor{\Pi}{_\mu_\nu},
\end{equation} 
in which the vector $q_{\mu}$ represents the heat flux, and $\tensor{\Pi}{_\mu_\nu} = \Pi \tensor{h}{_\mu_\nu} + \tensor{\pi}{_\mu_\nu}$, in which $\Pi$ is the viscous bulk pressure, $\tensor{h}{_\mu_\nu} = v_{\mu}v_{\nu} - \tensor{g}{_\mu_\nu}$ the projector and $\tensor{\pi}{_\mu_\nu}$ the anisotropic stress tensor. These quantities further satisfy $q_{\mu} v^{\mu} = \tensor{g}{^\mu^\nu} \tensor{\pi}{_\mu_\nu} = v^{\mu} \tensor{\pi}{_\mu_\nu} = 0$. It should also be noted that these objects are related via the first law of thermodynamics $d(\rho/\rho_{m}) = \mathfrak{T} d\mathfrak{s} - p d(1/\rho_{m})$. Hence, for a constant entropy density $\mathfrak{s}$ or a constant rest energy density $\rho_{m},$ it is obtained that
\begin{equation}
\label{eq22}    
\rho_{m} d\rho = (\rho + p) d\rho_{m} \ \ \text{and} \ \ \rho_{m}^{-1} d\rho = \mathfrak{T} d\mathfrak{s},
\end{equation}
in which $\rho$ is the total energy density, $p$ the isotropic pressure and $\mathfrak{T}$ the temperature.

The generalization of the approach of \cite{Mendoza2023} to Lyra manifolds consists in assuming that the matter Lagrangian is a function of $\phi$ and $\tensor{g}{_\mu_\nu}$ through the rest energy density and the entropy density. The source $\Omega$ is found from
\begin{equation}
\label{eq23}
\frac{\delta \mathcal{L}_{m}}{\delta \zeta^{a}} = \frac{\delta \rho_{m}}{\delta \zeta^{a}} \frac{\partial \mathcal{L}_{m}}{\partial \rho_{m}} + \frac{\delta \mathfrak{s}}{\delta \zeta^{a}} \frac{\partial \mathcal{L}_{m}}{\partial \mathfrak{s}},
\end{equation}
in which $\zeta^{a} = \phi$ or $\zeta^{a} = \tensor{g}{^\mu^\nu}$. It is first necessary to discover the expression for $\mathcal{L}_{m}$ through the functional derivatives of $\rho_{m}$ and $\mathfrak{s}$ with respect to the metric. Then, $\Omega$ is found by considering the resulting $\mathcal{L}_{m}$ and through the functional derivatives of the relevant thermodynamical quantities with respect to the Lyra field.

Without loss of generality, it is utilized the Eckart frame, in which the continuity equation is written as the one for a perfect fluid \cite{Mendoza2021}, namely, $\nabla_{\mu} (\rho_{m} v^{\mu}) = 0$, which by using the definition of the Lyra covariant derivative in  \eqref{eq7} yields
\begin{equation}
\label{eq24}
\frac{1}{\phi^{4} \sqrt{-g}} \frac{\partial (\phi^{3} \sqrt{-g} \rho_{m} v^{\mu})}{\partial x^{\mu}} = 0.
\end{equation}
As a result, $\delta (\phi^{3} \sqrt{-g} \rho_{m} v^{\mu}) = 0$. If $\delta (\tensor{g}{_\mu_\nu} v^{\mu} v^{\nu}) = 0$ is also used, since $v^{\mu} v_{\mu} = 1$, by performing the variation with respect to both $\phi$ and $\tensor{g}{^\mu^\nu}$, it follows that the functional derivatives of the rest energy density are given by
\begin{equation}
\label{eq25}    
\frac{\delta \rho_{m}}{\delta \tensor{g}{^\mu^\nu}} = - \frac{1}{2} \rho_{m} \tensor{h}{_\mu_\nu} \ \ \text{and} \ \ \frac{\delta \rho_{m}}{\delta \phi} = - \frac{3}{\phi}\rho_{m}.
\end{equation}

Regarding the derivatives of the entropy density, it is necessary to consider the equation for the energy-momentum tensor conservation generalized to Lyra manifolds, which is simply given by
\begin{equation}
\label{eq26}
\nabla^{\mu} \tensor{T}{_\mu_\nu} = 0.
\end{equation}
Contracting this expression with the four-velocity $v_{\nu}$, considering the Lyra geodesic equation $v^{\mu} \nabla_{\mu} v^{\nu} = 0$ and that $v_{\nu} \nabla_{\mu} v^{\nu} = 0$ yield
\begin{equation}
\label{eq27}
\xi v^{\mu} \nabla_{\mu} \mathfrak{s} = \nabla_{\mu} q^{\mu} + v_{\nu} v^{\mu} \nabla_{\mu} q^{\nu} + \Pi \nabla_{\mu} v^{\mu} + v_{\nu} \nabla_{\mu} \tensor{\pi}{^\mu^\nu},
\end{equation}
such that $\xi = - \rho_{m} \mathfrak{T}$. To find this result, we have used $\nabla_{\mu} (\omega v^{\mu}) - v^{\mu} \nabla_{\mu} p = - \xi v^{\mu} \nabla_{\mu} \mathfrak{s}$, since the relativistic enthalpy is given by $\omega = \rho + p = \rho_{m} (\mu + \mathfrak{T} \mathfrak{s})$, due to $d\mu = \rho_{m}^{-1} dp - \mathfrak{s} d\mathfrak{T}$ \cite{Mendoza2023}. Through this expression, by using the relations in \cite{Mendoza2023} and since $v_{\mu} \partial v^{\mu}/\partial \phi = v_{\mu} \partial q^{\mu}/\partial \phi = 0$ due to $v^{\mu} v_{\mu} = 1$ and $v_{\mu} q^{\mu} = 0$, is then possible to find the sought relations:
\begin{equation}
\label{eq28} 
\xi \frac{\delta \mathfrak{s}}{\delta \tensor{g}{^\mu^\nu}} = q_{(\mu}v_{\nu)} + \frac{1}{2} \tensor{\Pi}{_\mu_\nu} \ \ \text{and} \ \ \xi \frac{\delta \mathfrak{s}}{\delta \phi} = \frac{3}{\phi} \Pi.
\end{equation}

Therefore, using the first expressions of \eqref{eq25} and \eqref{eq28} into the energy-momentum tensor definition $\tensor{T}{_\mu_\nu} = -2 \delta \mathcal{L}_{m}/\delta \tensor{g}{^\mu^\nu} + \mathcal{L}_{m} \tensor{g}{_\mu_\nu}$ \cite{Mendoza2023} and considering the thermodynamical relations in \eqref{eq22}, it is straightforward to find that the matter Lagrangian for an imperfect fluid on the Lyra geometry is given by
\begin{equation}
\label{eq29}
\mathcal{L}_{m} = \rho.
\end{equation}
As a consequence, using this expression along with $\delta \rho_{m}/\delta \phi$ and $\delta \mathfrak{s}/\delta \phi$ of \eqref{eq25} and \eqref{eq28}, via the assumption \eqref{eq23}, into the definition \eqref{eq15}, leads to $-\Omega = \rho - 3(p + \Pi)$. However, since this is exactly the trace of \eqref{eq20}, it is thus found that the Lyra field source is proportional to the trace of the energy-momentum tensor:
\begin{equation}
\label{eq30}
\Omega = - T.
\end{equation}
This result was obtained without using a specific gravitational Lagrangian, and is valid for any matter field obeying the thermodynamical relations employed in this section. This is so because any energy-momentum tensor can be written as in Eq.\eqref{eq20} via its irreducible decomposition \cite{Rezzolla2013}.

\section{The Brans-Dicke-Gauss-Bonnet Lyra theory}
\label{sec6} 

The Brans-Dicke theory \cite{Brans1961} was one of the first scalar-tensor theories to be constructed and its solutions are well-explored in the literature \cite{Faraoni2004, Clifton2012}. One of its main contributions was to foster observational tests of gravitational phenomena, since for years it was one of the few competitors against GR \cite{Brans2005}. In its original Lagrangian density, it is only considered a scalar field non-minimally coupled to the Ricci scalar and a kinetic term with a constant coupling parameter divided by the scalar field \cite{Brans1961}. In later iterations, this coupling is generalized to a function of the scalar field \cite{Nordtvedt1970} and a scalar potential is considered \cite{Clifton2012, Goenner2012}. To construct a Lyra version of the generalized Brans-Dicke theory it is thus considered that
\begin{equation}
\label{eq50}
\mathcal{L}_{BDL} = \phi R - \frac{\omega(\phi)}{2} \nabla_{\mu} \phi \nabla^{\mu} \phi - 2 V(\phi),
\end{equation}
in which $\phi$ is the Lyra scale field, $R$ is given by \eqref{eq13} and such that $\nabla_{\mu}$ is the Lyra covariant derivative. 

Moreover, as stated in the Introduction \ref{sec:1}, the Gauss-Bonnet topological invariant on pseudo-Riemannian manifolds only contributes to four-dimensional gravitational phenomena if alternative approaches are considered. The same situation arises in the Lyra geometry, given that its Gauss-Bonnet term:
\begin{equation}
\label{eq31}
\mathscr{G} = R^{2} - 4 \tensor{R}{^\mu^\nu} \tensor{R}{_\mu_\nu} + \tensor{R}{^\mu^\nu^\alpha^\beta} \tensor{R}{_\mu_\nu_\alpha_\beta},
\end{equation}
in which the Riemann tensor is given by Eq.\eqref{eq11.1}, becomes a total derivative in the variation of the action with respect to the metric tensor. In addition, variation in relation to the Lyra function also conspires to the vanishing of this object in the final equations. 

However, the coupling of this term with a function of the Lyra scale field makes possible the emergence of contributions in the field equations. Therefore, unlike the usual 4D Gauss-Bonnet scalar-tensor theories of Riemannian manifolds, there is now the possibility of constructing a theory with the Gauss-Bonnet invariant which is formed solely with geometrical fields. The Lyra function allows the definition of scalar-tensor theories in the spirit of the geometrization brought by General Relativity.

Due to this fact, and inspired by the Riemannian dilatonic scalar-tensor theories, from expressions \eqref{eq50} and \eqref{eq31} the following action in the Lyra geometry for the gravitational sector is proposed: 
\begin{equation}
\label{eq32}
S_{g}[\tensor{g}{_\mu_\nu}, \phi] = \frac{1}{2\kappa} \int_{\mathcal{M}} \mathcal{L}_{g}[\tensor{g}{_\mu_\nu}, \phi] \,\phi^4 \sqrt{-g} \,d^{4}x,
\end{equation}
where the Brans-Dicke-Gauss-Bonnet Lyra Lagrangian is given by
\begin{equation}
\label{eq33}
\mathcal{L}_{g} = \eta(\phi) R + \alpha f(\phi) \mathscr{G} - \frac{\omega(\phi)}{2} \nabla_{\mu} \phi \nabla^{\mu} \phi - 2 V(\phi),
\end{equation}
in which the constant $\alpha$ has units of $[\kappa]^{-1}$ and $f(\phi)$ and $\eta(\phi)$ are dimensionless non-minimal coupling functions. As in Eq.\eqref{eq50}, $\omega(\phi)$ is the dimensionless kinetic coupling function and $V(\phi)$ is the autointeraction potential. It should be noted that the curvature invariants include terms containing $\phi$. The Ricci scalar is given by Eq.\eqref{eq13} and the Gauss-Bonnet Lyra scalar \eqref{eq31} can be written as
\begin{eqnarray}
\label{eq34}    
\mathscr{G} &=& \frac{1}{\phi^{4}} \mathscr{G}_{R} + \frac{4}{\phi^{4}} \mathcal{R} \nabla_{\mu} \phi \nabla^{\mu} \phi - \frac{4}{\phi^{3}} \mathcal{R} \Box \phi + \frac{8}{\phi^{3}} \tensor{\mathcal{R}}{_\mu_\nu} \nabla^{\mu} \nabla^{\nu} \phi \nonumber \\ 
&&- \frac{8}{\phi^{2}} (\nabla_{\mu} \nabla_{\nu} \phi) (\nabla^{\mu} \nabla^{\nu} \phi) + \frac{24}{\phi^{4}} (\nabla_{\mu} \phi \nabla^{\mu} \phi)^{2} \nonumber \\ 
&&+ \frac{8}{\phi^{2}} (\Box \phi)^{2} - \frac{24}{\phi^{3}} (\Box \phi) (\nabla_{\mu} \phi \nabla^{\mu} \phi),
\end{eqnarray}
in which $\mathscr{G}_{R}$ is the Riemannian Gauss-Bonnet term defined with respect to the Christoffel symbol of the metric $\tensor{g}{_\mu_\nu}$ and such that the covariant derivative is given by Eq.\eqref{eq7}. Naturally, the Lagrangian \eqref{eq50} is obtained if $\eta(\phi) = \phi$ and $\alpha = 0$. If $\eta(\phi) = 1$ in Eq.\eqref{eq33}, the result is a Lyra generalization of Einstein-Gauss-Bonnet scalar-tensor theories. If in addition $\alpha = \omega(\phi) = V(\phi) = 0$, it leads then to the LyST theory \cite{Cuzinatto2021}. 

To find the field equations that follow from the action \eqref{eq32} plus the matter action $S_{m}[\tensor{g}{_\mu_\nu}, \phi, \psi_{a}]$, where $\psi_{a}$ represents the matter fields, it is important to use the Bianchi identity of Lyra torsionless metric manifolds:
\begin{equation}
\label{eq36}  
\nabla_{\beta} \tensor{R}{_\alpha_\mu_\nu_\gamma} + \nabla_{\nu} \tensor{R}{_\alpha_\mu_\gamma_\beta} + \nabla_{\gamma} \tensor{R}{_\alpha_\mu_\beta_\nu} = 0,
\end{equation}
which leads to the usual contracted forms, but associated with Lyra covariant derivatives. The reason is that Eq.\eqref{eq36} simplifies the final equations so that only second order terms in the derivatives of the metric are displayed. 

Therefore, by ignoring the total Lyra covariant derivative terms, the variation of the action $S = S_{g} + S_{m}$ with relation to the metric yields
\begin{equation}
\label{eq37}
\eta(\phi) \tensor{G}{_\mu_\nu} + \tensor{\Sigma}{_\mu_\nu} + \alpha \tensor{\Lambda}{_\mu_\nu} + \tensor{\Theta}{_\mu_\nu} + V(\phi) \tensor{g}{_\mu_\nu} = \kappa \tensor{T}{_\mu_\nu},
\end{equation}
such that, due to the contracted Lyra-Bianchi identities that arise from Eq.\eqref{eq36}, the new tensors are defined as
\begin{eqnarray}
\label{eq37.1}
\tensor{\Sigma}{_\mu_\nu} &=& (\tensor{g}{_\mu_\nu} \tensor{g}{^\alpha^\beta} - \tensor{\delta}{^\alpha_\mu} \tensor{\delta}{^\beta_\nu}) (\eta_{\phi} \nabla_{\alpha} \nabla_{\beta} \phi + \eta_{\phi \phi} \nabla_{\alpha} \phi \nabla_{\beta} \phi), ~~~~~\\ 
\label{eq38}
\tensor{\Lambda}{_\mu_\nu} &=& f \tensor{H}{_\mu_\nu} + 2\tensor{\Upsilon}{_\alpha_\mu_\nu_\beta} (f_{\phi} \nabla^{\alpha} \nabla^{\beta} \phi + f_{\phi \phi} \nabla^{\alpha} \phi \nabla^{\beta} \phi), ~~~~~\\
\label{eq39}
\tensor{\Theta}{_\mu_\nu} &=& - \frac{\omega (\phi)}{2} \nabla_{\mu} \phi \nabla_{\nu} \phi + \frac{\omega (\phi)}{4} \tensor{g}{_\mu_\nu} \nabla^{\alpha} \phi \nabla_{\alpha} \phi, 
\end{eqnarray}
with $f = f(\phi)$, $f_{\phi} = \partial f / \partial \phi$, $f_{\phi \phi} = \partial^{2} f / \partial \phi^{2}$ and
\begin{eqnarray}
\label{eq40}  
\tensor{\Upsilon}{_\alpha_\mu_\nu_\beta} 
&=& 4 \tensor{R}{_\alpha_{(\mu}} \tensor{g}{_{\nu)}_\beta} - 2 \tensor{R}{_\alpha_\mu_\nu_\beta} - 2 \tensor{g}{_\mu_\nu} \tensor{R}{_\alpha_\beta} \nonumber ~~~~~\\
&&- 2 \tensor{R}{_\mu_\nu} \tensor{g}{_\alpha_\beta} - R \tensor{g}{_\alpha_\mu} \tensor{g}{_\beta_\nu} + \tensor{g}{_\mu_\nu} R \tensor{g}{_\alpha_\beta}. 
\end{eqnarray}
The tensor $\tensor{H}{_\mu_\nu}$ in relation \eqref{eq38} is given by
\begin{eqnarray}
\label{eq41}
\tensor{H}{_\mu_\nu} &=& 2 R \tensor{R}{_\mu_\nu} - 4 \tensor{R}{_\mu_\alpha} \tensor{R}{_\nu^\alpha} + 4 \tensor{R}{_\alpha_\mu_\nu_\beta} \tensor{R}{^\alpha^\beta} \nonumber ~~~\\ 
&&+ 2 \tensor{R}{_\mu^\alpha^\beta^\gamma} \tensor{R}{_\nu_\alpha_\beta_\gamma} - \frac{1}{2} \tensor{g}{_\mu_\nu} \mathscr{G},
\end{eqnarray}
with the curvature objects defined by expressions \eqref{eq11.1}, \eqref{eq12}, \eqref{eq13} and \eqref{eq34}. Note that, due to the Bach-Lanczos identity \cite{Tian2016} generalized to Lyra manifolds, $H_{\mu\nu}$ vanishes. 

Varying the action with respect to the Lyra scalar field $\phi$, disregarding the surface terms and substituting the Lyra field source relation \eqref{eq30} into the resulting equation leads to 
\begin{eqnarray}
\label{eq42}
\bigg(\eta + \phi \frac{\eta_{\phi}}{2} \bigg) R - 3 \eta_{\phi} \Box \phi - 3 \eta_{\phi \phi} \nabla^{\alpha} \phi \nabla_{\alpha} \phi && \nonumber \\
- 2 \alpha \tensor{\Upsilon}{_\alpha_\beta} (f_{\phi} \nabla^{\alpha} \nabla^{\beta} \phi + f_{\phi \phi} \nabla^{\alpha} \phi \nabla^{\beta} \phi) && \nonumber \\
- \frac{\omega(\phi)}{2} \nabla^{\alpha} \phi \nabla_{\alpha} \phi - 4 V(\phi) + \frac{\alpha}{2}f_{\phi} \phi \mathscr{G} && \nonumber \\
+ \frac{\omega(\phi)}{2} \phi \Box \phi + \frac{\omega_{\phi}}{4} \phi \nabla^{\alpha} \phi \nabla_{\alpha} \phi - \phi V_{\phi} &=& - \kappa T,
\end{eqnarray} 
such that $\tensor{\Upsilon}{_\alpha_\beta} = R \tensor{g}{_\alpha_\beta} - 2 \tensor{R}{_\alpha_\beta}$. Note that for $\eta = 1$ and $\alpha = V = \omega = 0$ this equation reduces to the LyST equation \eqref{eq15.1}. On the other hand, contracting Eq.\eqref{eq37} with $\tensor{g}{^\mu^\nu}$ yields
\begin{eqnarray}
\label{eq43}
- \eta R + 3 \eta_{\phi} \Box \phi + 3 \eta_{\phi \phi} \nabla^{\alpha} \phi \nabla_{\alpha} \phi && \nonumber \\
+ 2 \alpha \tensor{\Upsilon}{_\alpha_\beta} (f_{\phi} \nabla^{\alpha} \nabla^{\beta} \phi + f_{\phi \phi} \nabla^{\alpha} \phi \nabla^{\beta} \phi) && \nonumber \\
+ \frac{\omega(\phi)}{2} \nabla^{\alpha} \phi \nabla_{\alpha} \phi + 4 V(\phi) &=& \kappa T.
\end{eqnarray}  
As a consequence, a simpler equation for the scale field is found by the sum of the trace equation \eqref{eq43} with Eq.\eqref{eq42}, which can be written as
\begin{equation}
\label{eq44}
\Box \phi = -\frac{1}{\omega(\phi)}\bigg(\eta_{\phi} R + \alpha f_{\phi} \mathscr{G} + \frac{\omega_{\phi}}{2} \nabla^{\alpha} \phi \nabla_{\alpha} \phi - 2 V_{\phi} \bigg).
\end{equation}
It is interesting to note that Eq.\eqref{eq44} is a condition for diffeomorphism invariance of the field equations in the Lyra sense, such that taking the covariant divergence of Eq.\eqref{eq37} leads to relation \eqref{eq26}. It is also worth noting that to obtain the Lyra version of the Brans-Dicke wave equation $\Box \phi \propto T$ it is necessary to use Eq.\eqref{eq43} into \eqref{eq44}. The expressions \eqref{eq37} and \eqref{eq44} are then the Brans-Dicke-Gauss-Bonnet Lyra (BDGBL) field equations.

\section{The Lyra transformation} 
\label{sec5}

To describe any physical phenomenon, in particular those related to mechanics, it is first necessary to consider a particular coordinate system. It is also essential to assign a system of units for the characterization of physical quantities. For dynamics, for instance, this is done for space-like distances, time-like intervals and for the mass of objects. Such a system is usually chosen to be globally fixed in a spacetime manifold, like the International System of Units (SI), in which, for example, a meter corresponds to the same physical length everywhere. However, it is also possible to choose a system which has instead a spacetime dependence, so that physical units change in scale from one point to another on a manifold \cite{Faraoni2004, Faraoni2007, Quiros2000, Quiros2012}. 

In this sense, as pointed out by Dicke \cite{Dicke1962}, before a spacetime geometry is defined, a choice for the unit of length is required. The length unit could then be chosen to vary according to a non-null dimensionless scalar $\phi(x^{\mu})$. As a result, if one were to use another system which is controlled by a function $\psi(x^{\mu})$ or a globally fixed unit system, Physics would remain unaltered by such changes, since the choice of the system of units is arbitrary \cite{Dicke1962, Faraoni2004, Quiros2000, Quiros2012}. In the case of geometrical theories, this means that under such transformations field equations must covariantly transform, the line element has to be invariant and free particles need to follow geodesics.

Despite the freedom in choosing the system of units, the metric compatible pseudo-Riemannian manifolds commonly used in Physics are not adapted to transformations between local systems of units, since the use of a global system is primarily considered \cite{Quiros2000, Quiros2012}. The line element of Lorentzian geometry, for example, is invariant under coordinate transformations and global transformations of length units. This is a result of the fact that a global change of length unit can be written as a coordinate transformation $x^{\mu} \rightarrow x'^{\mu} = a x^{\mu}$ in which $a$ is a constant, so that $ds'^{2} = ds^{2}$ \cite{Wheeler2018}.

To solve this issue, a transformation to a system calibrated by an arbitrary dimensionless scalar can be performed in Riemannian geometry. Following a similar reasoning by \cite{Dicke1962, Faraoni2004}, since the coordinate element $dx^{\mu}$ is associated to a globally fixed system of units, performing a transformation to a local system leads to
\begin{equation}
\label{eq30.1}
dx^{\mu} \rightarrow d \Tilde{x}^{\mu} = \phi(x^{\mu}) dx^{\mu},
\end{equation}
such that $d \Tilde{x}^{\mu}$ is the new coordinate infinitesimal measuring ``rod" defined in a frame for which the length unit is point-dependent.\,The new line element is thus given by
\begin{equation}
\label{eq30.2}
d \Tilde{s}^{2} = \tensor{g}{_\mu_\nu} d \Tilde{x}^{\mu} d \Tilde{x}^{\nu},
\end{equation} 
which, using Eq.\eqref{eq30.1} yields
\begin{equation}
\label{eq30.3}
d \Tilde{s}^{2} = \tensor{\Tilde{g}}{_\mu_\nu} dx^{\mu} dx^{\nu} = ds^{2},
\end{equation} 
such that the line element $ds^{2}$ with metric components $\tensor{\Tilde{g}}{_\mu_\nu} = \phi^{2} \tensor{g}{_\mu_\nu}$ is defined in the global unit frame of $dx^{\mu}$. Note also that the notation used here is based on \cite{Faraoni2004, Faraoni2007}.

Therefore, the line element with components $\tensor{g}{_\mu_\nu}$ expressed in a frame of a locally varying unit of length is equal to the line element of a global length unit frame with metric components $\tensor{\Tilde{g}}{_\mu_\nu} = \phi^{2} \tensor{g}{_\mu_\nu}$. Both frames are then physically equivalent, the difference being in the interpretation. This consideration is evident for instance in the cosmological context. The components of the flat Friedmann-Lemaître-Robertson-Walker metric can be written with conformal time $\eta$ as $\tensor{\Tilde{g}}{_\mu_\nu} = a^{2}(\eta) \tensor{\eta}{_\mu_\nu}$, so that by Eq.\eqref{eq30.2} it follows that
\begin{equation}
\label{eq30.4}
d \Tilde{s}^{2} = d \Tilde{\eta}^{2} - d \Tilde{x}^{2} - d \Tilde{y}^{2} - d \Tilde{z}^{2}.
\end{equation}
As a consequence, an expanding universe is equivalently interpreted as a Minkowskian metric for which the unit of length has a decrement with conformal time governed by $\phi = a(\eta)$ \cite{Faraoni2004, Deruelle2011, Chiba2013}. 

Note that in order to write a spacetime interval with respect to a locally varying length unit, as in \eqref{eq30.2}, it is implemented the change given in Eq.\eqref{eq30.1}, since the coordinate elements act as infinitesimal measuring rods, carrying then the length unit information. Nonetheless, there are systems for which certain coordinates have no units or units different from that of length, so that some of the metric components are thus the objects who carry the units \cite{Mana2021}. For example, this is the case of the angular coordinates in the Schwarzschild metric in Droste coordinates. However, as pointed out in \cite{Mana2021}, the important aspect to focus on is the intrinsic units of a tensor, which is determined by the physical and operational meaning of the tensor. In this sense, what is truly important is the components and bases units combination, since the scalar $\phi$ modifies the resulting units of the bases and components combined. As a result, the essence is in the line element itself, since $d \Tilde{s}^{2} = ds^{2} = \phi^{2} \tensor{g}{_\mu_\nu} dx^{\mu} dx^{\nu}$ expresses exactly the fact that physical distances and time intervals can be equivalently written with respect to a global system of units ($d s^{2}$) or a spacetime-dependent one ($d \Tilde{s}^{2}$).

Therefore, although in Eq.\eqref{eq30.1} it is considered that the coordinate elements possess all the units information, using Eq.\eqref{eq30.1} is a general approach which simply manifests the arbitrariness in the use of physical units via the main expression \eqref{eq30.3} \cite{Dicke1962, Faraoni2004}. The interesting aspect of such assumption is that Eq.\eqref{eq30.1} is exactly the definition of the Lyra dual vector basis given by Eq.\eqref{eq3.1}, so that $\theta^{\mu} = d \Tilde{x}^{\mu}$. In the Lyra case, Eq.\eqref{eq3.1} is a consequence of Eq.\eqref{eq1}. However, assuming first Eq.\eqref{eq3.1} also naturally leads to the Lyra geometry formalism presented in section \ref{sec2}. The Lyra line element \eqref{eq5} can then be interpreted as being defined in a system with a spacetime-dependent length unit, that is, $d s^{2} = \tensor{g}{_\mu_\nu} \theta^{\mu} \theta^{\nu}$, so that $d s^{2}$ in the Lyra notation is equal to $d \Tilde{s}^{2}$ in the notation from Eq.\eqref{eq30.2}.

As a consequence, adapting Riemannian manifolds to include a point-dependent unit of length leads exactly to the Lyra geometry, as it can be observed by the affine and metric structures generated by the choice of the basis in Eq.\eqref{eq1}. This follows from the fact that the definition in Eq.\eqref{eq1} is equivalent to assume Eq.\eqref{eq30.1}, since it naturally implies in Eq.\eqref{eq3.1} via the orthonormality condition. Therefore, a Lyra reference system $(x^{\mu}, \phi)$ is a pair which comprises a coordinate system and a spacetime-dependent function that sets the unit of length.

From this perspective it becomes clear that a Lyra transformation $(x^{\mu}, \phi) \rightarrow (\bar{x}^{\mu}, \bar{\phi})$ is both a change in coordinates and a transformation between different local systems of units of length. It is important to note that the choice of the length unit is arbitrary by construction, so that the new function $\bar{\phi}$ is an arbitrary scalar which is not necessarily related to the $\phi$ defined in another frame. As a result, a Lyra tensor as defined in Eq.\eqref{eq3.2} being invariant under Lyra transformations means that such a tensor is invariant under changes in coordinates and length units, i.e. $\bar{T} = T$. For this to happen, the tensor components transformation law is then given by Eq.\eqref{eq4} and the bases must change according to Eq.\eqref{eq2}. Thus, as mentioned before, the length unit transformation acts on the whole tensor \eqref{eq3.2}, not solely on the units of its components \cite{Mana2021}.

It is important to note that a change in the unit of length can be performed independently of a change in coordinates \cite{Lyra1951, Dicke1962}. It is possible to change the unit system while keeping the coordinates fixed. Thus, for example, a Lyra transformation such as $(x^{\mu}, \phi) \rightarrow (\bar{x}^{\mu} = x^{\mu}, \bar{\phi} = 1)$ is a change from a generic Lyra frame to a global unit system, i.e.~a Riemannian frame. In this scenario, Eq.\eqref{eq4} leads to the following frame transformation law for the tensor components:
\begin{equation}
\label{eq30.5}
\hspace*{-0.1cm}
\bar{T}^{\mu_{1}...\mu_{k}}_{\ \ \ \ \ \ \ \ \nu_{1}...\nu_{l}} = \phi^{l-k} T^{\mu_{1}...\mu_{k}}_{\ \ \ \ \ \ \ \ \nu_{1}...\nu_{l}}.
\end{equation}
Therefore, as expected from Eq.\eqref{eq30.3}, the metric components transform as 
\begin{equation}
\label{eq30.6}
\tensor{\bar{g}}{_\mu_\nu} = \phi^{2} \tensor{g}{_\mu_\nu}.
\end{equation}
The line element in the rigid unit frame is then $d \bar{s}^{2} = \tensor{\bar{g}}{_\mu_\nu} d x^{\mu} d x^{\nu}$, such that $d \bar{s}^{2}$ in the Lyra formalism notation is equal to $d s^{2}$ in the notation from Eq.\eqref{eq30.3}.

To avoid confusion, from now on the Lyra geometry notation will be the only one used. Therefore, all symbols with an overbar represent mathematical objects defined in the global unit system, that is, with $\bar{\phi} = 1$. Accordingly, any symbol lacking an overscript refers then to a Lyra frame. So, for example, $d s^{2}$ refers to a variable length unit frame controlled by $\phi$ and $d \bar{s}^{2}$ to a rigid one. 

Regarding the curvature tensors, from Eq.\eqref{eq11.1} it is evident that the Riemann tensor defined in the global unit frame is simply given by $\tensor{\bar{R}}{^\rho_\mu_\gamma_\alpha}(\tensor{\bar{g}}{_\mu_\nu}, \bar{\phi} = 1) = \tensor{\bar{\mathcal{R}}}{^\rho_\mu_\gamma_\alpha}$. As a result, from the transformation law \eqref{eq30.5}, it is obtained 
\begin{equation}
\label{eq30.7}
\tensor{R}{^\rho_\mu_\gamma_\alpha} = \phi^{-2} \tensor{\bar{\mathcal{R}}}{^\rho_\mu_\gamma_\alpha}.
\end{equation}
Note that, due to Eq.\eqref{eq30.5}, the various forms of the Riemann tensor transform as follows: $\tensor{R}{^\rho^\mu^\gamma^\alpha} = \phi^{4} \tensor{\bar{\mathcal{R}}}{^\rho^\mu^\gamma^\alpha}$, $\tensor{R}{_\rho_\mu_\gamma_\alpha} = \phi^{-4} \tensor{\bar{\mathcal{R}}}{_\rho_\mu_\gamma_\alpha}$, $\tensor{R}{^\rho^\mu^\gamma_\alpha} = \phi^{2} \tensor{\bar{\mathcal{R}}}{^\rho^\mu^\gamma_\alpha}$ and $\tensor{R}{^\rho^\mu_\gamma_\alpha} = \tensor{\bar{\mathcal{R}}}{^\rho^\mu_\gamma_\alpha}$. As for its contractions, the transformed Ricci tensor is simply $\tensor{R}{_\mu_\nu} = \phi^{-2} \tensor{\bar{\mathcal{R}}}{_\mu_\nu}$ and the Ricci scalar is $R = \bar{\mathcal{R}}$. As a consequence, the Einstein tensor on a Lyra frame, i.e. $\tensor{G}{_\mu_\nu} = \tensor{R}{_\mu_\nu} - \frac{1}{2} R \tensor{g}{_\mu_\nu}$, is related to the Riemannian Einstein tensor of the metric $\tensor{\bar{g}}{_\mu_\nu}$ via
\begin{equation}
\label{eq30.8}
\tensor{G}{_\mu_\nu} = \phi^{-2} \tensor{\bar{\mathcal{G}}}{_\mu_\nu}.
\end{equation}
Observe that, as defined in section \ref{sec3}, the calligraphic letters refer to tensors defined as in Riemannian manifolds and with respect to the Levi-Civita connection. 

To discover how the Lyra connection \eqref{eq11} transforms, however, it is necessary to use Eq.\eqref{eq30.6}. Since $\tensor{g}{^\mu^\nu} = \phi^{2} \tensor{\bar{g}}{^\mu^\nu}$ from Eq.\eqref{eq30.5}, it follows that the transformed Lyra connection is given by
\begin{equation}
\label{eq30.9}
   \Gamma^{\alpha}_{\mu\nu} = \frac{1}{\phi} \bar{\bigg\{\genfrac{}{}{0pt}{0}{\alpha}{\mu \nu}\bigg\}} - \frac{1}{\phi} \delta^{\alpha}_{\mu} \nabla_\nu \phi,
\end{equation}
in which the Christoffel symbol with an overbar is defined with respect to $\tensor{\bar{g}}{_\mu_\nu}$. As a consequence, substituting expression \eqref{eq30.9} into the Lyra autoparallel equation \eqref{eq8} leads exactly to the Riemannian geodesic equation. That is, the transformed geodesic equation will possess only the Levi-Civita connection with respect to the metric $\tensor{\bar{g}}{_\mu_\nu}$. As for the case of a general frame transformation, it is obtained that $\phi \Gamma^{\alpha}_{\mu\nu} + \tensor{\delta}{^\alpha_\mu} \nabla_{\nu} \phi = \bar{\phi} \bar{\Gamma}^{\alpha}_{\mu\nu} + \tensor{\delta}{^\alpha_\mu} \bar{\nabla}_{\nu} \bar{\phi}$. 

If the conformal transformation $\tensor{g}{_\mu_\nu} = \phi^{-2} \tensor{\bar{g}}{_\mu_\nu}$ is instead first applied in the Lyra formalism, without referring to Eq.\eqref{eq30.5}, the relations \eqref{eq30.7}, \eqref{eq30.8} and \eqref{eq30.9} are also obtained. Such change leads first to Eq.\eqref{eq30.9}, which is an integrable projective transformation of the Christoffel symbol \cite{Olmo2022} with an additional rescaling by the Lyra field. As a consequence, \eqref{eq30.7}, \eqref{eq30.8} and the other Riemannian curvature tensors are naturally obtained. This result is somewhat expected due to the similarities between the Lyra and Weyl geometries, e.g.~Lyra's connection \eqref{eq11} without the antisymmetric part is equal to the WIST one \cite{Sen_1971, Lyra1951, Romero_2019, Cuzinatto2021}. This is the case because performing a conformal transformation in a Riemannian manifold leads directly to a WIST geometry \cite{Quiros2012, Quiros2018, Romero_2019}. The main difference, however, as mentioned in the Introduction \ref{sec:1}, is that the Lyra transformation acts on all tensors irrespective of the metric structure. That is, as expected, Eq.\eqref{eq30.5} is more general than solely considering Eq.\eqref{eq30.6}.

As a matter of fact, it is well known since Dicke that a conformal transformation can be interpreted as a local rescaling of units \cite{Dicke1962, Bekenstein1993, Quiros2000, Faraoni2004, Bisabr2004, Catena2007, Scholz2015}, so that a physical theory must also be conformal invariant \cite{Brans1961, Dicke1962, Bekenstein1980, Quiros2000, Faraoni2004, Faraoni2007, Scholz2017, Burikham2023}. In this sense, many authors since Eddington \cite{Canuto1977} have also pointed out that Weyl manifolds, in special, Weyl integrable spacetimes, are suitable for treating invariance under point-dependent transformations of units \cite{Ross1972, Quiros2000, Romero2012, Scholz2015, Yuan2015, Scholz2017, Wheeler2018, Quiros2018, Scholz2020, Burikham2023}. As for this work, it is shown that the Lyra geometry is a natural formalism to deal with such running units systems, since the Lyra transformation makes explicit the invariance of tensors under spacetime-dependent length units transformations. In section \ref{sec7}, it will be shown how the Lyra invariance is related to the Jordan and Einstein frame conundrum.

\subsection{The relation between LyST theory and General Relativity} 
\label{sec5.1} 

The core idea within the Lyra Scalar-Tensor Theory, as mentioned in section \ref{sec3}, is to generalize General Relativity to Lyra manifolds while promoting the Lyra function to a field. Nonetheless, since the Lyra geometry is shown here to be a formalism which deals with the invariance of spacetime intervals and other physical quantities under local length units systems transformations, the relation with GR is even more straightforward. The LyST field equation \eqref{eq14} can simply be written as $\tensor{G}{_\mu_\nu} = \kappa \tensor{T}{_\mu_\nu}$, so that changing to a global frame $\bar{\phi} = 1$ leads exactly to GR field equations: 
\begin{equation}
\label{eq30.10}
\tensor{\bar{\mathcal{G}}}{_\mu_\nu} = \kappa \tensor{\bar{\mathcal{T}}}{_\mu_\nu}.
\end{equation}
This is a result of the transformation law \eqref{eq30.5}, as it leads to Eq.\eqref{eq30.8} and 
\begin{equation}
\label{eq30.11}
\tensor{T}{_\mu_\nu} = \phi^{-2} \tensor{\bar{\mathcal{T}}}{_\mu_\nu}.
\end{equation}
The Riemannian energy conservation is also obtained, since using Eq.\eqref{eq30.5} in relation \eqref{eq26} yields
\begin{equation}
\label{eq30.12}
\nabla^{\mu} \tensor{T}{_\mu_\nu} = \phi^{-1} \bar{\nabla}^{\mu} \tensor{\bar{\mathcal{T}}}{_\mu_\nu} = 0,
\end{equation}
in which $\bar{\nabla}^{\mu}$ is the Riemannian covariant derivative with respect to the Levi-Civita connection. Thus, taking into account that Riemannian geodesics are also recovered, the LyST theory is simply a generalization of GR which covers covariance under transformations between dynamical length units systems. 

If a conformal transformation of the metric is solely considered first, to recover GR it is then necessary further assumptions. For the energy-momentum tensor, it is necessary to consider its form via the irreducible decomposition, since this object can be written in this manner for any fluid \cite{Rezzolla2013}. Thus, in natural units, a generic $\tensor{T}{_\mu_\nu}$ is given by Eq.\eqref{eq20}. By applying then Eq.\eqref{eq30.6}, if it is assumed that $\rho(x) = \bar{\rho}(x)$, $p(x) = \bar{p}(x)$ and $v_{\mu} = \phi^{-1} \bar{v}_{\mu}$, the perfect fluid part will be that of Riemannian manifolds but rescaled by $\phi^{-2}$. Considering also $q_{\mu} = \phi^{-1} \bar{q}_{\mu}$ and $\tensor{\Pi}{_\mu_\nu} = \phi^{-2} \tensor{\bar{\Pi}}{_\mu_\nu}$ leads then to Eq.\eqref{eq30.11}, in which $\tensor{\bar{\mathcal{T}}}{_\mu_\nu}$ is the usual Riemannian energy-momentum tensor. Therefore, since Eq.\eqref{eq30.6} leads to Eq.\eqref{eq30.8}, GR is recovered for such assumptions. Using then Eq.\eqref{eq30.9} and Eq.\eqref{eq30.11} into \eqref{eq26} leads also to \eqref{eq30.12}. In this sense, the LyST theory is just the conformal covariant version of GR \cite{Quiros2000, Deruelle2011, Wheeler2018}.

As for the second LyST equation \eqref{eq15.1}, since it is just the trace of \eqref{eq14}, it is obtained that $\bar{\mathcal{R}} = - \kappa \bar{\mathcal{T}}$ in both of the above approaches. Nonetheless, it should be noted that this equation does not bring any new information to the field equations, since it can be solely found from the first one \eqref{eq14}. That is, since the trace equation is simply a sum of the ten equations in Eq.\eqref{eq14}, a solution satisfying \eqref{eq14} will automatically satisfy the trace equation. Moreover, if the action \eqref{eq14.1} is not a functional of $\phi$, the resulting theory will be the same, since variation with respect to the metric leads naturally to \eqref{eq14}, for which taking the trace leads then to $R = - \kappa T$. Therefore, the Lyra function cannot be rigorously viewed as a dynamical field in the original LyST case, since the result of its variation yields no new equation. 

In essence, the local unit system is to be regarded on equal footing with the coordinate system, so that a LyST metric solution is defined on a predetermined pair $(x^{\mu}, \phi)$. However, in practice, high-symmetry considerations are always necessary to solve the field equations, so that some of the metric components are fixed, making it possible to determine an equation for the Lyra function. Therefore, it is possible to do an exchange of degrees of freedom by fixing a metric component and using instead the Lyra scalar. As an example, in the case of Eq.\eqref{eq30.4}, the FLRW solution can be written as $\tensor{\bar{g}}{_\mu_\nu} = a^{2} (\eta) \tensor{\eta}{_\mu_\nu}$ on a global unit frame or such that $\tensor{g}{_\mu_\nu} = \tensor{\eta}{_\mu_\nu}$ and $\phi = a(\eta)$, so that the only change is on the physical interpretation. A similar approach is adopted in \cite{Cuzinatto2021, Sobrero2024}, in which the spherical symmetry enables a consistent equation for $\phi$. Note, however, that such approaches were considered before this work, i.e.~before the Lyra function being directly identified with a local lenght unit system, so that now it is clear that it is always possible to transform to another length unit system controlled by some scalar function.

Furthermore, the approach used in \cite{Cuzinatto2021} is not always guaranteed to generate a valid equation, since in the LyST scenario $\phi$ cannot be properly seen as a field. Thus, as mentioned before, a more rigorous approach is to predefine the form of the Lyra function or to transform later the solutions to a frame of interest. In the cosmological context, for example, a scale factor solution can be found first using GR and then such scalar can be used as the new function which controls the unit of length, so that the resulting structure will be that of the LyST theory. Therefore, any GR solution can then have a varying unit LyST representation for which the unit frame is locally controlled by a scalar of interest.

\subsection{The relation of Brans-Dicke-Gauss-Bonnet Lyra theory with its Riemannian version} \label{subsec6.1} 
 
In the previous subsection, it is shown that the LyST theory becomes General Relativity after a particular Lyra transformation is applied. It is interesting, thus, to employ the same procedure to the Brans-Dicke-Gauss-Bonnet Lyra field equations to find the corresponding theory in the Riemannian frame associated with the metric $\tensor{\bar{g}}{_\mu_\nu}$. For this purpose, it is necessary to note that the derivatives of the Lyra scale function change according to the Lyra transformation \eqref{eq30.5}. That is, from $\nabla_{\mu} \phi = \partial_{\mu} \ln{\phi}$ and \eqref{eq30.5}, it follows that $\nabla_{\mu} \nabla_{\nu} \phi = \phi^{-2} \bar{\nabla}_{\mu} \bar{\nabla}_{\nu} \phi$, $\nabla^{\mu} \nabla^{\nu} \phi = \phi^{2} \bar{\nabla}^{\mu} \bar{\nabla}^{\nu} \phi$, $\Box \phi = \bar{\Box} \phi$, $\nabla^{\mu} \phi \nabla^{\nu} \phi = \phi^{2} \partial^{\mu} \phi \partial^{\nu} \phi$ and $\nabla_{\alpha} \phi \nabla^{\alpha} \phi = \partial_{\alpha} \phi \partial^{\alpha} \phi$, in which $\bar{\nabla}_{\mu}$ is exactly defined as in GR. Such results are equally obtained via relation \eqref{eq30.9}. Note also that the contraction of tensors in a particular Lyra frame is made with respect to the metric tensor components of that frame.

Due to the notation employed, it is easy to perceive that the equations \eqref{eq37} and \eqref{eq44} have the same external form of the generalized Brans-Dicke theory and Gauss-Bonnet gravity combination of Riemannian geometry \cite{Clifton2012, Tian2016}, with the main differences being the Lyra covariant derivatives, the Riemann tensor of Lyra manifolds and the $\phi$-dependent energy-momentum tensor. Therefore, considering the transformation law \eqref{eq30.5}, i.e., utilizing \eqref{eq30.6}, \eqref{eq30.7}, \eqref{eq30.8}, \eqref{eq30.9} and \eqref{eq30.11} into \eqref{eq37}, it is easily shown that
\begin{equation}
\label{eq30.13}
\eta(\phi) \tensor{\bar{\mathcal{G}}}{_\mu_\nu} + \tensor{\bar{\Sigma}}{_\mu_\nu} + \alpha \tensor{\bar{\Lambda}}{_\mu_\nu} + \tensor{\bar{\Theta}}{_\mu_\nu} + V(\phi) \tensor{\bar{g}}{_\mu_\nu} = \kappa \tensor{\bar{\mathcal{T}}}{_\mu_\nu},
\end{equation}
in which
\begin{eqnarray}
\label{eq30.14}
\tensor{\bar{\Sigma}}{_\mu_\nu} &=& (\tensor{\bar{g}}{_\mu_\nu} \tensor{\bar{g}}{^\alpha^\beta} - \tensor{\delta}{^\alpha_\mu} \tensor{\delta}{^\beta_\nu}) (\eta_{\phi} \bar{\nabla}_{\alpha} \bar{\nabla}_{\beta} \phi + \eta_{\phi \phi} \partial_{\alpha} \phi \partial_{\beta} \phi), ~~~~~\\ 
\label{eq30.15}
\tensor{\bar{\Lambda}}{_\mu_\nu} &=& 2\tensor{\bar{\Upsilon}}{_\alpha_\mu_\nu_\beta} (f_{\phi} \bar{\nabla}^{\alpha} \bar{\nabla}^{\beta} \phi + f_{\phi \phi} \partial^{\alpha} \phi \partial^{\beta} \phi), ~~~~~\\
\label{eq30.16}
\tensor{\bar{\Theta}}{_\mu_\nu} &=& - \frac{\omega (\phi)}{2} \partial_{\mu} \phi \partial_{\nu} \phi + \frac{\omega (\phi)}{4} \tensor{\bar{g}}{_\mu_\nu} \tensor{\bar{g}}{^\alpha^\beta} \partial_{\alpha} \phi \partial_{\beta} \phi, 
\end{eqnarray}
such that $\tensor{\bar{H}}{_\mu_\nu} = 0$ and $\tensor{\bar{\Upsilon}}{_\alpha_\mu_\nu_\beta}$ is defined as in Eq.\eqref{eq40} but with relation to the new metric and its Riemannian objects: $\tensor{\bar{\mathcal{R}}}{_\alpha_\mu_\nu_\beta}$, $\tensor{\bar{\mathcal{R}}}{_\mu_\nu}$ and $\bar{\mathcal{R}}$. As for the scalar field equation, doing the same approach in \eqref{eq44} leads to
\begin{equation}
\label{eq30.17}
\bar{\Box} \phi = -\frac{1}{\omega(\phi)}\bigg(\eta_{\phi} \bar{\mathcal{R}} + \alpha f_{\phi} \bar{\mathscr{G}}_{R} + \frac{\omega_{\phi}}{2} \partial^{\alpha} \phi \partial_{\alpha} \phi - 2 V_{\phi} \bigg),
\end{equation} 
given that, due to Eq.\eqref{eq30.5}, $\mathscr{G} = \bar{\mathscr{G}}_{R}$, in which $\bar{\mathscr{G}}_{R}$ is the Gauss-Bonnet term defined in the torsionless and metric-compatible Riemannian manifold $\tensor{\bar{g}}{_\mu_\nu}$. 

Therefore, as expected from the correspondence of the Lyra geometry with local length unit systems explored in this section, the BDGBL equations in the global unit frame, i.e.~Eq.\eqref{eq30.13} and Eq.\eqref{eq30.17}, are exactly the combination of the generalized Brans-Dicke gravity with Gauss-Bonnet scalar-tensor theories of Riemannian manifolds \cite{Brans1961, Clifton2012, Tian2016, Sotiriou2007}. Naturally, as mentioned in subsection \ref{sec5.1}, this correspondence also extends to the energy-momentum tensor conservation and to the geodesic equation, since they also reduce to their Riemannian counterparts in this unit system ($\bar{\phi} = 1$), so that the corresponding connection is the Levi-Civita one. 

As a consequence, since equations \eqref{eq30.13} and \eqref{eq30.17} are related to \eqref{eq37} and \eqref{eq44} via a local length unit transformation, the Lyra geometry enables a direct four-dimensional geometrization of the Riemannian scalar-tensor equations \eqref{eq30.13} and \eqref{eq30.17}, since transforming to the Lyra frame provides a geometrical meaning to the scalar field $\phi$. In special, note that for $\eta(\phi) = 1$ it is obtained the Einstein-Gauss-Bonnet scalar-tensor theories of Riemannian geometry \cite{Tian2016, Sotiriou2007, Nojiri2005a}, so that the Lyra formalism enables then the 4D geometrization of commonly known theories with the Gauss-Bonnet invariant.

As for the particular case of $\alpha = 0$ and $\eta(\phi) = \phi$, note that the equations reduce to
\begin{eqnarray}
\label{eq51.1}
\phi \tensor{\bar{\mathcal{G}}}{_\mu_\nu} - \bar{\nabla}_{\mu} \bar{\nabla}_{\nu} \phi + \tensor{\bar{g}}{_\mu_\nu} \bar{\Box} \phi - \frac{\omega(\phi)}{2} \partial_{\mu} \phi \partial_{\nu} \phi && \nonumber ~~~~~~~~~~~~~\\
+ \frac{\omega(\phi)}{4} \tensor{\bar{g}}{_\mu_\nu} \partial_{\alpha} \phi \partial^{\alpha} \phi + \tensor{\bar{g}}{_\mu_\nu} V(\phi) &=& \kappa \tensor{\bar{\mathcal{T}}}{_\mu_\nu},
\end{eqnarray}
and
\begin{equation}
\label{eq54.1}
\omega(\phi) \bar{\Box} \phi + \bar{\mathcal{R}} + \frac{\omega_{\phi}}{2} \partial_{\alpha} \phi \partial^{\alpha} \phi - 2 V_{\phi} = 0,
\end{equation}
which are then the generalized Brans-Dicke field equations \cite{Brans1961, Clifton2012}. A similar equivalence, as briefly mentioned in the Introduction, was partially found by \cite{Almeida2014} in the Weyl geometry, but restricted to vacuum solutions and not encompassing inertial motion. In addition, reference \cite{Soleng1988} has found that \eqref{eq51.1} with $\tensor{\bar{\mathcal{T}}}{_\mu_\nu} = 0$ can be obtained from the Lyra geometry, although their geometrical considerations are very different than the ones assumed here. 

Brans-Dicke gravity can then be physically reinterpreted as a completely geometrical theory via the Lyra geometry. This approach further strengthens the physical concept that was considered in the inception of this theory \cite{Brans1961}, that is, Dicke's view on the Mach's principle. The Mach-Dicke conjecture states that the coupling between spacetime and matter must change according to the distribution of mass in the Universe \cite{Brans2005}. The Lyra framework allows a natural implementation of such idea, since the mathematical objects directly responsible for defining spacetime intervals in the Lyra frame, i.e.~$\tensor{g}{_\mu_\nu}$ and $\phi$, can be the gravitational fields. The length unit in which the line element is written is locally determined by the function $\phi(x^{\mu})$, which can then be defined from the matter distribution of the Universe through a particular Lagrangian. Hence, in such case, the coupling between matter and spacetime has the simple geometrical meaning of determining the unit of length. 

Nonetheless, it should be noted that in the Riemannian Brans-Dicke theory the varying gravitational coupling is given by the scalar field alone. In the Lyra case, however, the Lyra field changes the coupling as $\kappa_{eff} = \kappa/\phi$, since this field has no physical units due to its role in setting the length unit. Furthermore, also owing to this same definition and to the base definition \eqref{eq1}, $\phi$ cannot be zero by construction, since this would mean an ill-defined length unit. This fact does not affect known solutions, such as cosmological ones or general asymptotically flat spacetimes \cite{Brans1961, Clifton2012}, since $\phi = 0$ naturally leads to unphysical scenarios, as it would imply in a gravitational coupling that diverges. Nevertheless, the limit $\phi \rightarrow 0$ is still possible via this framework.

It is important to note that there are two slightly different approaches which lead to the same results. It is possible to start from a particular scalar-tensor Lagrangian in the Riemannian geometry and transform, through \eqref{eq30.5}, to a system with the length unit controlled by the scalar field itself. Or, as done in subsection \ref{sec6}, start from a generic Lyra frame and define a particular dynamics for the objects which determine spacetime intervals, that is, a Lyra invariant Lagrangian relating $\tensor{g}{_\mu_\nu}$ and $\phi$ to matter fields. In this case, the general relation \eqref{eq30} is an important result, since it shows that both approaches are consistent, leading to the same field equations in an independent manner (see Eq.\eqref{eq42}).  

Therefore, both ways provide identical results, which is consistent with the field equations \eqref{eq37} and \eqref{eq44} being just a different representation of \eqref{eq30.13} and \eqref{eq30.17}, so that the only difference is in the use of distinct systems for the unit of length. This is also observed at the level of the action \eqref{eq32}, since transforming to a global unit system leads to $\sqrt{-g} = \phi^{-4} \sqrt{-\bar{g}}$ \cite{Cuzinatto2021} and  
\begin{equation}
\label{eq30.18}
\bar{\mathcal{L}}_{g} = \eta(\phi) \bar{\mathcal{R}} + \alpha f(\phi) \bar{\mathscr{G}}_{R} - \frac{\omega(\phi)}{2} \tensor{\bar{g}}{^\mu^\nu} \partial_{\mu} \phi \partial_{\nu} \phi - 2 V(\phi),
\end{equation}
which leads then to Eq.\eqref{eq30.13} and Eq.\eqref{eq30.17}. Naturally, the inverse is also true, as going back to the running unit system leads back to \eqref{eq33}. Note that, due to the Lyra transformation law \eqref{eq30.5}, $\mathcal{L}_{g} = \bar{\mathcal{L}}_{g}$.

Nevertheless, some minor procedural differences arise by considering first the Lyra frame. The scalar equation \eqref{eq44} is obtained from the sum of the trace equation with the expression that comes from the variation with respect to the Lyra scale field. However, if one begins in a Riemannian frame and then transform, the same equation is instead obtained solely from the variation of the action with respect to the scalar field. Furthermore, by analyzing the passage of the BDGBL equations from the Lyra to the Riemannian frame or vice-versa, it also becomes clear why the LyST theory cannot be properly considered a scalar-tensor theory. While for the BDGBL theory a dynamical scalar is still present after transforming to a global unit system, for the LyST equations this does not occur, since the metric is the only field in the Riemannian frame. At a fundamental level, the most important aspect is that the Lyra approach provides a geometrical interpretation to the scalar field, as in the Lyra frame the scalar gravitational field is one of the foundational constructs used to define the ``measuring rods and clocks".

\section{The Jordan and Einstein frames on Lyra manifolds} 
\label{sec7}   

Soon after the first Brans-Dicke papers \cite{Brans1961, Brans1961b} were published, Dicke wrote a subsequent work exploring a peculiarity of his theory. In \cite{Dicke1962}, he showed that the Brans-Dicke equations can be rewritten as General Relativity sourced by an additional scalar matter field through a coordinate-dependent transformation of the units of measure. Such transformation is shown to be conformal, but since it is only a local redefinition of units, this new form is simply shown to be a different representation of the same physical theory. Nonetheless, the notation utilized in \cite{Dicke1962} is a bit misleading, since it does not make explicit the invariance of objects through such transformations, for instance, the geodesic equation is altered by the scalar field after such change.

As a result, combined with the fact that \cite{Dicke1962} has not received due attention over the years, a lot of discussions regarding the physical equivalence of theories that differ only by a conformal transformation have emerged in the literature \cite{Faraoni2004, Quiros2012, Quiros2019}. Known as the Jordan and Einstein frame conundrum \cite{Quiros2019}, in which the Einstein frame is the one for which the Brans-Dicke equations can be written as GR, such debate, as discussed in \cite{Faraoni2004}, mainly arises if one does not properly take into account that the Einstein frame is defined in a locally varying system of units. To address such issue, this section is thus devoted to show that the Jordan and Einstein frame relation is made more explicit by utilizing Lyra manifolds.

It is important to point out the notation employed here, which is based on the Lyra formalism. For instance, a physical scalar quantity $A$ associated to distance measurements can be written as $A = \{A\} [L]$, in which $\{A\}$ is a pure dimensionless numerical quantity and $[L]$ is the length unit. As a local length unit is related to a rigid unit system by $[L] = \phi [\bar{L}]$, transforming to a global unit frame means that $\{A\} = \phi^{-1} \{\bar{A}\}$, since a physical quantity must be invariant by such arbitrary transformation: $A = \bar{A}$ (with $\bar{A} = \{\bar{A}\} [\bar{L}]$). Thus, the pure numerical quantity has the inverse transformation of the length unit, so that in this notation it is clear that a physical quantity is invariant to a change in the unit of length. 

Nonetheless, this is not the notation mainly utilized in \cite{Dicke1962} and in current works \cite{Deruelle2011, Chiba2013, Faraoni2007}. The invariance of the line element, for example, is not explicit in the notation utilized by Dicke. As a result, as it can be seen in \cite{Deruelle2011}, the invariance of a particle's action under such transformation requires that the unit of mass be also transformed \cite{Dicke1962, Faraoni2004}. In the Lyra formalism, however, one can solely perform a spacetime-dependent length unit transformation, such that physical quantities are explicitly invariant under such change. In a Lyra frame, for instance, the action of a charged particle in a electromagnetic field is
\begin{equation}
\label{eq30.19}
S_{p} = - \int m c \sqrt{\phi^{2} \tensor{g}{_\mu_\nu} dx^{\mu} dx^{\nu}} + q \int A_{\mu} v^{\mu} d\tau,
\end{equation}
in which $v^{\mu}$ is given by Eq.\eqref{eq3}. Transforming to a global unit system, by using Eq.\eqref{eq30.6} and Eq.\eqref{eq30.5}, leads then to $S_{p} = \bar{S}_{p}$, in which
\begin{equation}
\label{eq30.20}
\bar{S}_{p} = - \int m c\,d\bar{s} + q \int \bar{A}_{\mu} \bar{v}^{\mu} d\tau.
\end{equation}
Therefore, due to the explicit invariance of the line element ($ds^{2} = d \bar{s}^{2}$), a mass rescaling is not necessary, so that $m = \bar{m}$ as expected from the Lyra notation. Moreover, note that $A_{\mu} = \phi^{-1} \bar{A}_{\mu}$, $\bar{v}^{\mu} = dx^{\mu} / d\tau$ and $d\tau = d\bar{\tau}$. Note also that Eq.\eqref{eq30.19} leads to the Lorentz equation written with respect to a running unit system $m v^{\alpha} \nabla_{\alpha} v^{\mu} = q \tensor{F}{^\mu_\beta} v^{\beta}$, where the derivative is the Lyra one and such that $\tensor{F}{_\mu_\nu} = \nabla_{\mu} A_{\nu} - \nabla_{\nu} A_{\mu}$ \cite{Sobrero2024}.

To study the change between the Jordan and Einstein frame from the perspective of Lyra transformations, it is considered the Brans-Dicke Lagrangian in the global length unit system $\bar{\phi} = 1$, i.e.~the Jordan frame. Denoting the scalar field as $\psi(x^{\mu})$, the action is simply given by
\begin{equation}
\hspace*{-0.1cm}
\label{eq30.21}
\bar{S}_{g} = \frac{1}{2\kappa} \int \bigg( \psi \bar{\mathcal{R}} - \frac{\omega(\psi)}{\psi} \partial_{\mu} \psi \partial^{\mu} \psi - U(\psi) \bigg) \sqrt{- \bar{g}}\,d^{4}x,
\end{equation}
in which $\bar{\mathcal{R}}$ is the usual Ricci scalar with respect to the Levi-Civita connection. For the Einstein frame, it is chosen a Lyra transformation that leads to a length unit controlled by $\phi = 1/\sqrt{\psi}$, since Eq.\eqref{eq30.6} becomes $\tensor{g}{_\mu_\nu} = \psi \tensor{\bar{g}}{_\mu_\nu}$ \cite{Chiba2013}. Using thus the transformation law \eqref{eq30.5} into Eq.\eqref{eq30.21} yields
\begin{equation}
\hspace*{-0.29cm}
\label{eq30.22}
S_{g} = \frac{1}{2\kappa} \int \bigg( \psi R - \frac{\omega(\psi)}{\psi} \nabla_{\mu} \psi \nabla^{\mu} \psi - U(\psi) \bigg) \frac{\sqrt{- g}}{\psi^{2}}\,d^{4}x,
\end{equation}
where 
\begin{equation}
\label{eq30.23}
R = \psi \mathcal{R} + \frac{3}{\psi} \Box \psi - \frac{3}{2 \psi^{2}} \nabla_{\mu} \psi \nabla^{\mu} \psi 
\end{equation}
due to Eq.\eqref{eq13} and such that $\nabla_{\mu} \psi = \sqrt{\psi} \,\partial_{\mu}\psi$. Note also that $\bar{\mathcal{R}} = R$, $\bar{\omega}(\psi) = \omega(\psi)$ and $\bar{U}(\psi) = U(\psi)$. Eliminating the total derivative term by using Eq.\eqref{eq14.3} and defining the new variable $\zeta$ through
\begin{equation}
\label{eq30.24}
\left(\frac{d \zeta}{d \psi}\right)^{2} = \frac{2 \omega(\psi) + 3}{2 \kappa \psi^{2}},
\end{equation} 
leads then to the action
\begin{equation}
\label{eq30.25}
S_{g} = \int \bigg( \frac{\mathcal{R}}{2 \kappa} - \frac{1}{2} \partial_{\mu} \zeta \partial^{\mu} \zeta - V(\zeta) \bigg) \sqrt{- g}\,d^{4}x,
\end{equation}
which looks like GR sourced by a dynamical scalar field and such that $V(\zeta) = U(\psi)/2 \kappa \psi^{2}$.

Although the action \eqref{eq30.25} closely resembles General Relativity with a scalar source, it is important to note that $\mathcal{R}$ does not correspond to the curvature scalar of the varying length unit system in which the action itself is written. The scalar \eqref{eq30.23} is the correct Ricci-Lyra invariant which measures the scalar curvature of the manifold of Eq.\eqref{eq30.25}. In addition, $\psi^{-2}\sqrt{- g}\,d^{4}x$ is the true volume element of this case and $\partial_{\mu} \zeta$ is not covariant in this frame, since its Lyra covariant derivative is defined as $\nabla_{\mu} \zeta = \sqrt{\psi} \,\partial_{\mu} \zeta$. Therefore, despite the resulting action \eqref{eq30.25} being widely known in the literature \cite{Faraoni2004, Chiba2013, Clifton2012}, it is usually obtained solely through a conformal transformation of the metric components, and not from a well-defined structure like the Lyra geometry. Thus, as shown here, the Lyra formalism reveals that Eq.\eqref{eq30.25} is not properly written with objects that are covariant with respect to the local length unit frame utilized.

For such reasons, specially due to the use of Eq.\eqref{eq14.3}, if one wishes to follow a more consistent approach in obtaining the field equations with matter sources, it is necessary to do the variation of the action written in the form of Eq.\eqref{eq30.22}. In this scenario, from Eq.\eqref{eq30.5}, note that a Riemannian matter action simply transform as $\bar{S}_{m}(\tensor{\bar{g}}{_\mu_\nu}, ...) = S_{m}(\psi^{-1} \tensor{g}{_\mu_\nu}, ...)$, in which any matter field also transform according to Eq.\eqref{eq30.5}. Variation of Eq.\eqref{eq30.22} plus $S_{m}$ with respect to $\tensor{g}{_\mu_\nu}$ leads thus to
\vspace*{-0.2cm}
\begin{eqnarray}
\label{eq30.26}
\psi \tensor{G}{_\mu_\nu} - \nabla_{\mu} \nabla_{\nu} \psi + \tensor{g}{_\mu_\nu} \Box \psi - \frac{\omega(\psi)}{\psi} \nabla_{\mu} \psi \nabla_{\nu} \psi && \nonumber ~~~~~~~~~~~~~\\
+ \frac{\omega(\psi)}{2 \psi} \tensor{g}{_\mu_\nu} \nabla_{\alpha} \psi \nabla^{\alpha} \psi + \frac{1}{2} \tensor{g}{_\mu_\nu} U(\psi) &=& \kappa \tensor{T}{_\mu_\nu},
\end{eqnarray}
such that by using \eqref{eq12}, \eqref{eq30.23} and \eqref{eq30.24} it yields
\begin{equation}
\label{eq30.27}
\tensor{\mathcal{G}}{_\mu_\nu} = \kappa \bigg(\frac{\tensor{T}{_\mu_\nu}}{\psi^{2}} + \partial_{\mu} \zeta \partial_{\nu} \zeta - \frac{1}{2} \tensor{g}{_\mu_\nu} \partial_{\alpha} \zeta \partial^{\alpha} \zeta - \tensor{g}{_\mu_\nu} V(\zeta) \bigg).
\end{equation}
As for the scalar equation, variation of the action with respect to $\psi$ or $\phi$ leads to
\begin{equation}
\label{eq30.28}
(3 + 2\omega) \Box \psi + \omega_{\psi} \nabla_{\alpha} \psi \nabla^{\alpha} \psi + 2 U - \psi U_{\psi} = - \kappa T,
\end{equation}
which by using relation \eqref{eq30.24} results in
\begin{equation}
\label{eq30.29}
\Box \zeta + \frac{d \psi}{d \zeta} \partial_{\mu} \zeta \partial^{\mu} \zeta - \psi \frac{d V}{d \zeta} = - \frac{1}{2} \frac{d \psi}{d \zeta} \frac{T}{\psi^{2}}.
\end{equation}
It is important to emphasize that the Laplace-Beltrami operator in the equation above is defined with the Lyra covariant derivative, so that it is given by
\begin{equation}
\label{eq30.30}
\Box \zeta = \frac{1}{\phi^{4} \sqrt{-g}} \frac{\partial (\phi^{2} \sqrt{-g}\,\partial^{\mu} \zeta)}{\partial x^{\mu}}.
\end{equation}
If a Riemannian-like operator is then defined as
\begin{equation}
\label{eq30.31}
\widehat{\Box} \zeta \coloneqq \frac{1}{\sqrt{-g}} \frac{\partial (\sqrt{-g}\,\partial^{\mu} \zeta)}{\partial x^{\mu}},
\end{equation}
it is possible to obtain that 
\begin{equation}
\label{eq30.32}
\widehat{\Box} \zeta - \frac{d V}{d \zeta} = - \frac{1}{2 \psi} \frac{d \psi}{d \zeta} \frac{T}{\psi^{2}},
\end{equation}
since $\psi \widehat{\Box} \zeta$ is equal to the first two terms of Eq.\eqref{eq30.29}.

The resulting equations \eqref{eq30.27} and \eqref{eq30.32} are equal to the usual Brans-Dicke equations in the Einstein frame form obtained throughout the literature \cite{Chiba2013, Faraoni2004, Clifton2012}, but with a crucial exception in the energy-momentum tensor notation. The energy-momentum tensor in Eq.\eqref{eq30.26} is in the Lyra frame, so that its definition is given by Eq.\eqref{eq14.2}. As expected, since a matter Lagrangian is a Lyra scalar $\bar{\mathcal{L}}_{m} = \mathcal{L}_{m}$, $\sqrt{-\bar{g}} = \psi^{-2} \sqrt{-g}$ and $\tensor{\bar{g}}{^\mu^\nu} = \psi \tensor{g}{^\mu^\nu}$, it follows that $\tensor{T}{_\mu_\nu}$ is related to the usual Riemannian energy-momentum tensor through Eq.\eqref{eq30.11}. As a result, the tensor $\bm{T} = \tensor{T}{_\mu_\nu} \theta^{\mu} \otimes \theta^{\nu}$ is invariant when transforming to a Riemannian frame. Moreover, note that to construct the $\tensor{T}{_\mu_\nu}$ for a fluid, it is necessary to follow the prescription explored in section \ref{sec4}, where objects are properly defined with respect to the Lyra frame utilized, such as using the Lyra four-velocity defined in Eq.\eqref{eq3}. 

Nonetheless, the usual energy-momentum tensor components defined in Einstein frame analyses does not correspond to Eq.\eqref{eq14.2}, so that the resulting tensor is not invariant under local length unit transformations. The conventional definition of the energy-momentum tensor in such case is given by \cite{Chiba2013, Faraoni2004}
\begin{equation}
\label{eq30.33}
\tensor{\widehat{T}}{_\mu_\nu} \coloneq - \frac{2}{\sqrt{-g}} \frac{\delta (\mathcal{L}_{m} \phi^{4} \sqrt{-g})}{\delta \tensor{g}{^\mu^\nu}},
\end{equation}
which is then related to the $\tensor{T}{_\mu_\nu}$ of the Lyra frame $\phi = 1/\sqrt{\psi}$ and to the Riemannian energy-momentum tensor through
\vspace{-0.2cm}
\begin{equation}
\label{eq30.34}
\tensor{\widehat{T}}{_\mu_\nu} = \frac{\tensor{T}{_\mu_\nu}}{\psi^{2}} = \frac{\tensor{\bar{\mathcal{T}}}{_\mu_\nu}}{\psi},
\end{equation}
such that $\widehat{T} \coloneq \tensor{g}{^\mu^\nu} \tensor{\widehat{T}}{_\mu_\nu} = T/\psi^{2} = \bar{\mathcal{T}}/\psi^{2}$. Thus, from the Einstein to the Jordan frame, the tensor $\bm{\widehat{T}} = \tensor{\widehat{T}}{_\mu_\nu} \theta^{\mu} \otimes \theta^{\nu}$ transform as $\bm{\widehat{T}} = \psi^{-2} \bar{\bm{\mathcal{T}}}$, whereas $\bm{T} = \bar{\bm{\mathcal{T}}}$ is invariant. 

As a consequence of definition \eqref{eq30.33}, the resulting conservation equation usually presented in the literature conveys then the impression that the energy-momentum tensor of the Einstein frame is not conserved in the sense of Eq.\eqref{eq26}. However, the reason for this is in the use of ill-defined objects, since using instead the Lyra covariant derivative \eqref{eq7} and Eq.\eqref{eq14.2} leads to the typical conservation equation \eqref{eq26}. This equation is then related to the one in the Riemannian frame through Eq.\eqref{eq30.12}. Naturally, the same occurs at the geodesic equation level, it might seem that free particles do not follow geodesics, but as seen in section \ref{sec5}, by using the Lyra formalism such equation is simply given by $v^{\mu} \nabla_{\mu} v^{\nu} = 0$. This is also observed in the Lorentz equation obtained from Eq.\eqref{eq30.19}. 

Therefore, by treating the Einstein frame as a Lyra frame, it becomes evident that geodesic motion and the conservation of the energy-momentum tensor are preserved. The Lyra geometry highlights that the difference between the Jordan and Einstein frames is only on the choice for the length unit system, so that the dynamics of particles and classical fields remains the same \cite{Faraoni2007}. It is also interesting to point out that the classical equivalence between both frames is verified in \cite{Chiba2013} through the use of cosmological observables, despite the typical notation that does not emphasize such correspondence. As a result, although equations \eqref{eq30.27} and \eqref{eq30.32} are easier to handle, it is more appropriate to consider the relations \eqref{eq30.26}, \eqref{eq30.28} and \eqref{eq26}, since they are written with respect to objects which are covariant with respect to local length unit transformations, i.e.~relations \eqref{eq7}, \eqref{eq12}, \eqref{eq30.23} and \eqref{eq14.2}.

\section{Final remarks}
\label{sec8} 

We have constructed a purely geometrical version of the generalized Brans-Dicke gravity on four-dimensional Lyra manifolds. The constructed theory also contains the Gauss-Bonnet invariant of the Lyra geometry, for which non-vanishing contributions of it appear in the four-dimensional field equations due to the non-minimal coupling with the Lyra scale function. To obtain the field equations, we have generalized an existing method to discover the matter Lagrangian of an imperfect fluid on Lyra manifolds, in which it was necessary to calculate the functional derivatives of the rest energy density and the entropy density in relation to the gravitational fields. Based on these results, it was found that the Lyra scale function source is equal to minus the trace of the energy-momentum tensor. This relation is valid as long as the components of the energy-momentum tensor in the irreducible decomposition framework satisfy the thermodynamical relations shown in Section \ref{sec4} (e.g.~perfect fluids, scalar fields, electromagnetic fields etc).

It was shown that adapting pseudo-Riemannian geometry to include a spacetime-dependent length unit system leads exactly to the Lyra geometry, such that the Lyra function is the object responsible for locally controlling the unit of length. We have shown that Lyra transformations comprise both coordinate and local transformations of length units. In the Lyra geometry it becomes evident that distances and other physical quantities must remain unaltered under an arbitrary change in the length unit utilized, since the line element and tensors are explicit invariant under Lyra transformations. The Lyra formalism highlights that local length unit systems must be treated on equal footing with the coordinates, so that as pointed out by Dicke in \cite{Brans1961}, ``the physical content of the
theory should be contained in the invariants of the group of position-dependent transformations of units and coordinate transformations.''

By transforming to a global length unit system, for which the Lyra function is constant, we have shown that Riemannian geometry is obtained and that the Lyra Scalar-Tensor theory becomes General Relativity. The LyST theory is shown to be just the GR generalization which covers covariance under point-dependent transformation of length units. We have also shown that the Brans-Dicke-Gauss-Bonnet Lyra theory constructed here is just its Riemannian scalar-tensor counterpart but written with respect to a frame for which the length unit is controlled by the scalar field itself. It is argued that this approach is more consistent with the Mach-Dicke principle, so that in the Lyra frame the scalar field has a geometrical meaning. As a result, the Lyra geometry can be seen as the natural setting for scalar-tensor theories, since there is always a frame for which the scalar which defines the length unit is a field of the theory, so that both the metric and the scalar field are the objects directly responsible for defining spacetime intervals.

The Lyra geometry further makes evident that the Jordan and Einstein frames are only different representations of the same classical physical theory, in which the Einstein frame is written with respect to a local length unit system. We have shown that the Einstein frame can be treated as a Lyra frame for which the scale function is the inverse of the square root of the Brans-Dicke scalar field. Due to its well-defined notation, it is shown that the energy-momentum tensor is conserved and that geodesic motion is preserved in the Einstein frame. Although the resulting equations are similar to GR with a scalar source, it is important to note that they are not written with objects which are covariant under local length units transformations. The appropriate form must be written with respect to Lyra covariant derivatives and the Ricci-Lyra tensor and scalar. We note that the energy-momentum tensor components definition commonly used in the literature is not covariant under a change between local length units systems, which can lead to numerous misunderstandings in the matter part of the usual Einstein frame equations. We note that the Einstein frame is just one of infinite possible frames in which the length unit is locally defined by some function of the scalar field.

Furthermore, it is important to remark that the Lyra structure is based on the consideration of a basis which is not the canonical one, that is, $\theta^{\mu} = \tensor{\xi}{^\mu_\alpha} dx^{\alpha}$ with $\tensor{\xi}{^\mu_\alpha} = \phi \tensor{\delta}{^\mu_\alpha}$, which corresponds then to the use of a local length unit system. We are currently working on a formalism which properly deals with disformal transformations, so that $\tensor{\xi}{^\mu_\alpha} = \psi \tensor{\delta}{^\mu_\alpha} + f(\psi) \partial^{\mu} \psi \partial_{\alpha} \psi$, which corresponds then to a length unit that also changes according to the gradient of $\psi$ \cite{Bekenstein1993}. We are specially interested in probing Horndeski gravity and Degenerate Higher-Order Scalar-Tensor (DHOST) theories in this framework \cite{Langlois2018b}. In addition, we are currently analysing the Palatini variational approach on Lyra manifolds. We are also studying cosmological linear perturbations to assess how matter perturbations behave in the Einstein frame from the perspective of the energy-momentum tensor which is covariant under Lyra transformations, since the frame independence of perturbations is not trivial \cite{Chiba2013, Francfort2019}. For future works, we wish to analyse if the equivalence between the Jordan and Einstein frame still holds at the quantum level by utilizing the Lyra geometry.

\vspace{0.4cm}
\section*{Acknowledgments}

E.C.V. and F.S. are grateful to professor Mario Novello, Nami Svaiter and the \textit{Pequenos Seminários} for the insightful discussions that enriched this work. We are also thankful to the valuable assistance of professor Martín Makler. We are deeply thankful for the helpful conversations with William Iania and Alexandre M. R. Almeida. This project was only made possible thanks to the financial support provided by CNPq, CAPES and FAPERJ. This study was financed in part by the Coordenação de Aperfeiçoamento de Pessoal de Nível Superior – Brasil (CAPES) – Finance Code 001. F.S. aknowledges the financial support from CNPq and FAPERJ (PhD merit fellowship - FAPERJ Nota 10, 203.709/2025). This preprint has not undergone peer review or any post-submission improvements or corrections. The Version of Record of this article is published in \textit{General Relativity and Gravitation}, and is available online at https://www.doi.org/10.1007/s10714-025-03509-8.

\bibliography{apssamp}

\end{document}